\documentclass[10pt,journal,compsoc]{IEEEtran}

\usepackage{amsmath}
\usepackage{amsfonts}
\usepackage{algorithm}  
\usepackage{algpseudocode}  
\usepackage{graphics}
\usepackage{textcomp}
\usepackage{amsmath} 
\usepackage{mathtools}
\usepackage{amssymb}  
\usepackage{color}
\usepackage{url}
\usepackage{commath}
\usepackage{caption}[font=small,labelfont=bf]
\usepackage{array} 
\usepackage{color} 
\usepackage{footnote}
\makesavenoteenv{tabular}
\makesavenoteenv{table}
\makesavenoteenv{figure}
\usepackage{enumitem}
\usepackage{siunitx}
\usepackage{tabularx}
\usepackage{subcaption}
\usepackage{lineno}
\usepackage{booktabs} 
\usepackage{array}
\usepackage{tikz}
\usepackage{booktabs}
\usepackage{multirow}
\usepackage{graphics}

\usepackage{graphicx}
\usepackage{adjustbox}
\usepackage{comment}
\usepackage{dblfloatfix}

\usepackage{hyperref}
\hypersetup{
    colorlinks=true,
    linkcolor=teal,
    filecolor=cyan,      
    urlcolor=teal,
    }
    
\usepackage{colortbl}
\usepackage{color}

\begin{document}

\title{MOCAS: A Multimodal Dataset for Objective Cognitive Workload Assessment on Simultaneous Tasks} 

\author{Wonse Jo$^{\dag}$, Ruiqi Wang$^{\dag}$, Go-Eum Cha, Su Sun, Revanth Krishna Senthilkumaran, \\Daniel Foti, and Byung-Cheol Min$^{*}$
\IEEEcompsocitemizethanks{\IEEEcompsocthanksitem W. Jo, R.Q. Wang, G. Cha, S. Sun, R.K. Senthilkumaran, and B.-C. Min are with SMART Lab, Department of Computer and Information Technology, Purdue University, West Lafayette, IN 47907, USA (e-mail: [jow, wang5357, cha20, sun931, senthilr, minb] @purdue.edu).
\IEEEcompsocthanksitem D. Foti is with Department of Psychological Sciences, Purdue University, West Lafayette, IN, USA (e-mail: foti@purdue.edu).
\IEEEcompsocthanksitem $\dag$: equal contribution
\IEEEcompsocthanksitem $*$: corresponding author}
\thanks{Manuscript received.}}

\markboth{IEEE Transactions on Affective Computing}%
{Shell \MakeLowercase{\textit{et al.}}: Bare Demo of IEEEtran.cls for Computer Society Journals}

%

\IEEEtitleabstractindextext{%
\begin{abstract}
This paper presents \textit{MOCAS}, a multimodal dataset dedicated for human cognitive workload (CWL) assessment. In contrast to existing datasets based on virtual game stimuli, the data in MOCAS was collected from realistic closed-circuit television (CCTV) monitoring tasks, increasing its applicability for real-world scenarios. To build \textit{MOCAS}, two off-the-shelf wearable sensors and one webcam were utilized to collect physiological signals and behavioral features from 21 human subjects. After each task, participants reported their CWL by completing the NASA-Task Load Index (NASA-TLX) and Instantaneous Self-Assessment (ISA). Personal background (e.g., personality and prior experience) was surveyed using demographic and Big Five Factor personality questionnaires, and two domains of subjective emotion information (i.e., arousal and valence) were obtained from the Self-Assessment Manikin (SAM), which could serve as potential indicators for improving CWL recognition performance. Technical validation was conducted to demonstrate that target CWL levels were elicited during simultaneous CCTV monitoring tasks; its results support the high quality of the collected multimodal signals.
\end{abstract}

}

\maketitle

\IEEEdisplaynontitleabstractindextext

%
\IEEEpeerreviewmaketitle

\section{Introduction}
\label{sec:introduction}
\IEEEPARstart{H}{umans} serve as the core part of any human-machine interaction system; thus, the cognitive workload of human operators is a critical concern in the design and implementation of such systems \cite{adams2002critical}. 
Cognitive workload (CWL) can be defined as the quantitative amount of mental loads exceeding the operator's ability to perform tasks \cite{wickens2002multiple}. It can also be considered as a mental strain to respond to tasks' demands for performing specific tasks \cite{cain2007review}.

Previous research \cite{marshall2002index,patten2006driver,patten2004using,just2003neuroindices,pomplun2003pupil} has repeatedly shown that awareness and perception of human CWL can improve the performance of a human-machine interaction system as such awareness allows identifying and alleviating task errors that result from situations of task over-load or under-load. Monitoring the CWL of human operators in a human-robot team, for instance, enables the workload and autonomy levels of the robots to be adjusted as needed to help human operators efficiently and productively maintain their working state \cite{mina2020adaptive}. Similarly, recognizing the CWL of human drivers can improve the safety of human-vehicle systems \cite{just2003neuroindices}.

\begin{figure*}[t]
    \centering
    \includegraphics[width=0.8\linewidth]{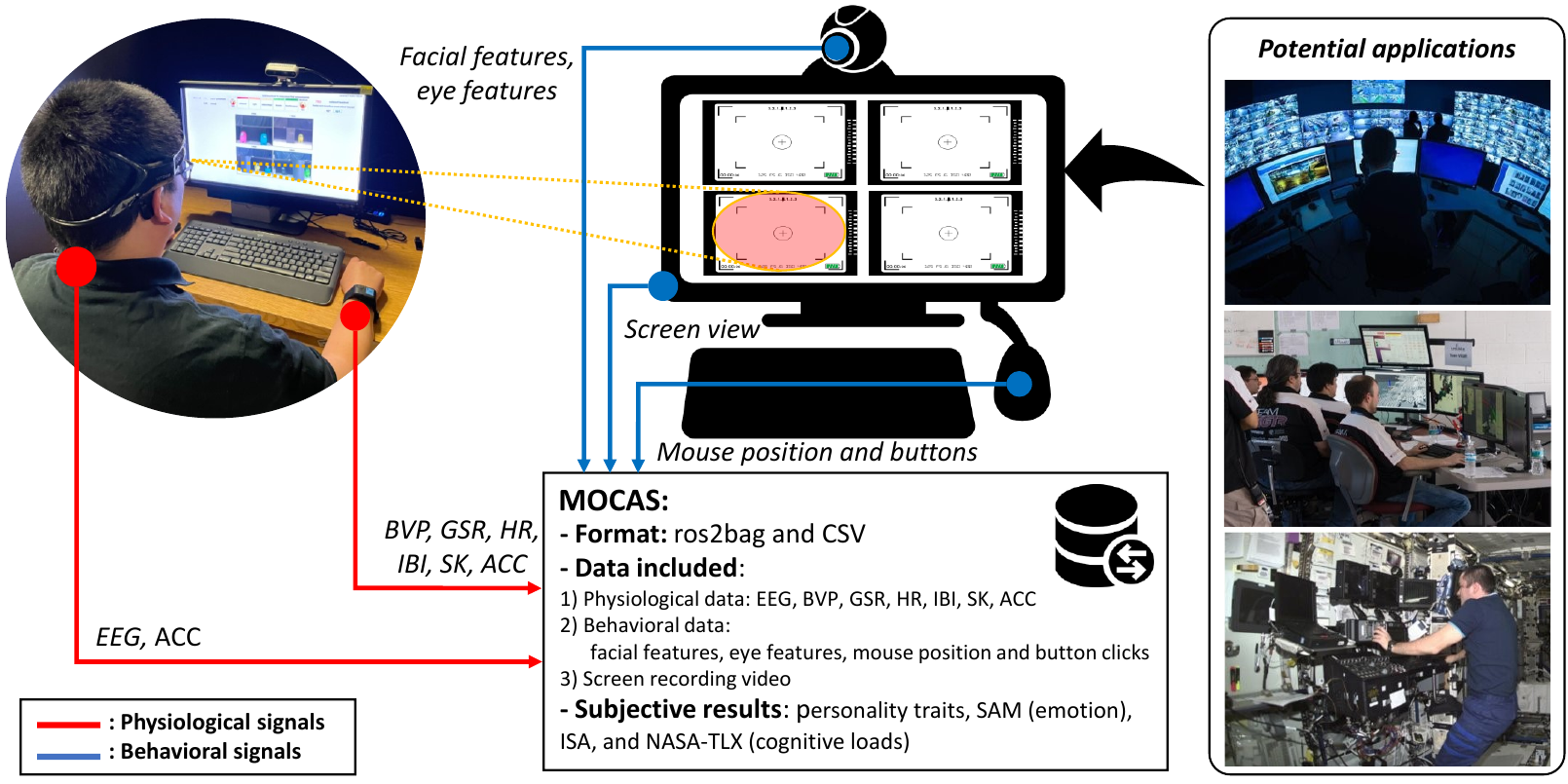}
    \caption{Illustration of the design of MOCAS dataset. MOCAS was designed with consideration of a task scenario typical in real-world human-machine systems, in which one human subject undertakes a simultaneous CCTV monitoring task and multiple sensors track that person’s physiological and behavioral metrics.} 
    \label{fig:scenario}
\end{figure*}

Generally, CWL assessments can be categorized as subjective or objective. Subjective CWL estimation focuses on the self-assessment of one's workload via subjective questionnaires such as the NASA-Task Load Index (NASA-TLX) \cite{hart1988development}. Objective measurement of CWL relies on quantitative data concerning one's physiological or behavioral responses when facing a certain level of workload. Accordingly, objective assessments can be further categorized as physiological or behavioral based on the data types utilized \cite{debie2019multimodal}. Physiological assessments perceive human biological metrics such as electroencephalography (EEG), functional near-infrared spectroscopy (fNIRS), skin temperature (SKT), and heart rate (HR), which change according to the reaction of the human nervous system to workload fluctuations. Behavioral assessments similarly analyze human behavioral responses such as facial landmarks, eye features, body gestures, and poses in relation to different levels of workload. 

However, none of the aforementioned signals are optimal in terms of accuracy, robustness, and rapidity due to the drawbacks inherent in utilizing an unimodal data source \cite{debie2019multimodal}. For example, different sensors have different sensitivity to different task scenarios and to different human subjects, therefore it is non-trivial to identify one single metric that is both efficient and accurate for general scenarios \cite{chen2013multimodal}. Moreover, noise and the failure of a single-modality sensor can lead to serious errors and even invalidity of the recognition system. Multimodal fusion-based CWL assessments have been proposed as a solution to this problem; such assessments combine bio-signals from multiple modalities to provide latent and important information that is unobtainable from a single modal source, and have achieved better performance \cite{debie2019multimodal,coffey2012measuring,putze2014hybrid,liu2017mental,liu2017multisubject,son2013identification,wang2022husformer}. Additionally, personality traits have been shown capable of influencing observed responses when different human subjects face the same level of workload; that is, under a given level of CWL, different personality traits can generate physiological and behavioral signals with different features \cite{gajos2017influence,gjoreski2020datasets,wigboldus2004capacity,choi2022immersion}. In addition, human emotion has been proven to have a close relation with CWL \cite{fraser2012emotion,van2009tuning,parsons2022interaction}; for instance, one's emotional state may directly affect CWL through expanding or shrinking cognitive resources \cite{plass2019four}. Thus, investigating personal traits, human emotion, and their potential relation with CWL can broaden our awareness of individual differences and suggest additional indicators for use in CWL assessments, thereby improving the accuracy and robustness of those assessments.

Even as multimodal fusion methods are being increasingly utilized in the area of objective CWL assessment, the availability of open-source and sizable multimodal datasets remains limited, making it difficult for researchers to ensure their models and algorithms are fairly reproducible and verifiable. Furthermore, existing open-source datasets \cite{lim2018stew,mcdonald2011quasar,albuquerque2020wauc,schneegass2013data,beh2021maus,gjoreski2020datasets,mijic2019mmod} were collected in the course of participants playing virtual games, such as dual N-back games \cite{jaeggi2008improving, owen2005n} and Multi-Attribute Task Battery (MATB) \cite{vyvey2018loaded, kalyuga2009evaluating}, which weakens the applicability of those datasets to real-world CWL recognition in human-machine interactions; these virtual games often feature artificial scenarios or simplified tasks, which may not adequately represent the complexity and dynamics found in real-world situations. Also, they usually focus on specific cognitive processes, such as working memory or attention, and may overlook the interplay between multiple cognitive processes, emotions, and contextual factors that typically arise in real-world contexts \cite{debie2019multimodal,jakobi1995noise}. In addition, according to our best knowledge, no extant dataset covers behavioral data and emotional information (annotations), and only one \cite{gjoreski2020datasets} includes the personality traits of the human subjects, but this is limited to considering the cognitive loads and performance for game-based tasks that mean it is not suitable on realistic applications. 

To fill the above-mentioned gaps, we constructed the \textbf{M}ultimodal Dataset for \textbf{O}bjective \textbf{C}ognitive Workload \textbf{A}ssessment on \textbf{S}imultaneous Tasks (MOCAS). To better mimic real-world scenarios, we designed and utilized a simultaneous closed-circuit television (CCTV) monitoring task to elicit target cognitive load, where participants interact with a physical multi-robot system in real-time and may face potential realistic dynamics and latency caused by the robotic system. The data in MOCAS is based upon 21 human subjects and consists of physiological data collected from two wearable sensors, behavioral data obtained from a camera, subjective CWL annotations via the Instantaneous Self-Assessment (ISA) \cite{jordan1992instantaneous} and NASA-TLX, subjective emotion annotations via the Self-Assessment Manikin (SAM) \cite{bradley1994measuring}, personal trait background surveyed from the Big Five Factor personality questionnaire \cite{jackson2002six}, and raw screen video recordings. MOCAS represents a useful addition to current research in the field of multimodal fusion CWL assessment: to our best knowledge, MOCAS is the first open-access dataset to obtain both physiological and behavioral data along with both CWL and emotion annotations including the personal traits and background of subjects. All data was collected from off-the-shelf and user-friendly sensors; accordingly, an assessment model built from our dataset can be applied to everyday applications with ease and efficiency.

\section{Dataset Design}
\subsection{Dataset Design}
MOCAS was designed with consideration of a task scenario typical in real-world human-machine systems (as illustrated in Fig. \ref{fig:scenario}), in which one human subject undertakes a simultaneous CCTV monitoring task and multiple sensors track that person's physiological and behavioral metrics. This dataset is intended to serve as an open-access, sizable, and multimodal data source for research in the field of objective assessment of cognitive workload, with the following aims:

\begin{itemize}
        \item  Support the study of cognitive workload recognition with a focus on real-world applications using human-machine systems.
        \item  Encourage research on improving the accuracy of cognitive workload assessment using real-world applications of multimodal fusion approaches, especially combining physiological and behavioral data.
        \item  Offer novel opportunities for investigating how awareness of a subject's personality traits and emotion states can improve cognitive workload assessment performance.
\end{itemize}

\begin{table*}[t]
    \caption{Comparison of the MOCAS dataset with the existing cognitive workload assessment datasets.}
    \label{tab:T1}
    \resizebox{\linewidth}{!}{%
    \begin{tabular}{cccccccc}
    \hline\hline
    \textbf{\begin{tabular}[c]{@{}c@{}}Name\\ (year)\end{tabular}} & \textbf{\begin{tabular}[c]{@{}c@{}}Number of\\ Participants\end{tabular} } & \textbf{Scenarios/Stimuli} & \textbf{\begin{tabular}[c]{@{}c@{}}Physiological\\ Data\end{tabular}}          & \textbf{\begin{tabular}[c]{@{}c@{}}Behavioral\\ Data\end{tabular}}                   & \textbf{\begin{tabular}[c]{@{}c@{}}Personal\\ Traits\end{tabular}} & \textbf{\begin{tabular}[c]{@{}c@{}}Cognitive\\ annotation\end{tabular}} & \textbf{\begin{tabular}[c]{@{}c@{}}Emotional\\ annotation\end{tabular}}
      \\ \hline 
    \begin{tabular}[c]{@{}c@{}}CSAC \cite{mcdonald2011quasar}\\ (2011)\end{tabular}     & 8                     & \begin{tabular}[c]{@{}c@{}}Multi-attribute \\ task battery (MATB)\end{tabular}                             & EEG, EOG                                                                       & X                                                                                    & X & O & X    \\ \hline
    
    \begin{tabular}[c]{@{}c@{}}STEW \cite{lim2018stew}\\ (2018)\end{tabular}     & 48                    & \begin{tabular}[c]{@{}c@{}}Single-session \\ simultaneous\\ capacity (SIMKAP)\end{tabular}                 & EEG                                                                            & X                                                                                    & X  &  O & X                  \\ \hline
    \begin{tabular}[c]{@{}c@{}}MMOD-COG \cite{mijic2019mmod}\\ (2019)\end{tabular} & 40                    & Reading and math test                                                                                      & ECG, GSR, HR                                                                   & X                                                                                    & X &  O & X                    \\ \hline
    \begin{tabular}[c]{@{}c@{}}WAUC \cite{albuquerque2020wauc}\\ (2020)\end{tabular}     & 48                    & \begin{tabular}[c]{@{}c@{}}Physical (Biking and treadmill) \\ and mental activities (MATB-II)\end{tabular} & \begin{tabular}[c]{@{}c@{}}EEG, ECG, ACC, TEMP,\\  GSR, BVP, RESP\end{tabular} & X                                                                                    & X  &   O & X               \\ \hline

    \begin{tabular}[c]{@{}c@{}}Cog Load \cite{gjoreski2020datasets}\\ (2020)\end{tabular} & 23                    & N-back game and video game                                                                                 & ACC, GSR, TEMP, RR                                                             & X                                                                                    & O & O & X                  \\ \hline
    \begin{tabular}[c]{@{}c@{}}Snake \cite{gjoreski2020datasets}\\ (2020)\end{tabular}    & 23                    & \begin{tabular}[c]{@{}c@{}}Smartphone game:\\  snake\end{tabular}                                          & ACC, GSR, TEMP, RR                                                             & X                                                                                    & O  & O & X                 \\ \hline
    \begin{tabular}[c]{@{}c@{}}CLAS \cite{markova2019clas}\\ (2021)\end{tabular}     & 62                    & \begin{tabular}[c]{@{}c@{}}Math and logic\\  problem\end{tabular}                                          & ECG, PPG, GSR, ACC                                                             & X                                                                                    & X & O & X                       \\ \hline
    \begin{tabular}[c]{@{}c@{}}Tufts \cite{huangfNIRS2MW2021}\\ (2021)\end{tabular}    & 68                    & N-back game                                                                                                & fNIRS                                                                          & X                                                                                   & X & O & X   \\ \hline
    
    \begin{tabular}[c]{@{}c@{}}MAUS \cite{beh2021maus}\\ (2021)\end{tabular}     & 22                    & N-back game                                                                                                & ECG, GSR, PPG                                                                  & X & X & O & X  \\ \hline
    \begin{tabular}[c]{@{}c@{}}MOCAS\\ (Ours, 2022)\end{tabular}   & 30 & Realisctic CCTV monitoring task                                                                                            & \begin{tabular}[c]{@{}c@{}}EEG, PPG, GSR, HR,\\  BVP, ACC, TEMP \end{tabular}         & \begin{tabular}[c]{@{}c@{}} Raw facial images, \\Facial features, \\ Eye features, \\ Mouse movement \\ \end{tabular} & O & O & O \\ \hline
    \end{tabular}
    }

    \footnotesize{ \vspace{4pt}''X" means not included, and ''O" means included.}
\end{table*}

Distinct from existing datasets related to cognitive workload, MOCAS evokes a target CWL with realistic CCTV monitoring tasks; it also includes more comprehensive multimodal data collected from both physiological and behavioral sensors, the personal backgrounds of participants (e.g., experience, preference, and personality traits), and emotion information (annotations). The major distinctions between existing CWL datasets and MOCAS are summarized in Table \ref{tab:T1}. 
Compared to other related datasets listed in Table \ref{tab:T1}, MOCAS stands out for its inclusion of multimodal data and features from physiological sensors and behavioral devices, including EEG, PPG, GSR, HR, ACC, and TEMP data, as well as facial video and monitor screen recordings. Additionally, MOCAS includes more diverse and realistic activities, which better reflect real-life scenarios. Furthermore, the dataset provides comprehensive annotations, such as labels for different activities, affective states, and cognitive load, that can be used to train and evaluate machine learning models for various applications, such as emotion recognition and mental workload estimation.

\subsection{Ethics statement}

This study and the building of MOCAS was reviewed and approved by the Purdue University Institutional Review Board (IRB) with approval number IRB-2021-1813. The informed consent form for participants included the purpose and procedure of data collection, specific types of data collected, possible risks or discomforts, compensation and benefits, protocols to protect privacy and confidentiality, and permission for potential usage of the collected data (publishing the dataset and conducting related research). Each participant was provided with the approved informed consent form upon their arrival at the experimental site for data collection, was asked to fully read and understand the content, and was asked to sign a written signature if s/he agreed to participate in the data collection procedure and approved the potential usage of the collected data. Participants also acknowledged that they had the right to terminate the data collection at any time they wanted.

All collected raw data, consent forms, and questionnaire forms (on which each participant was assigned and presented with a participant number) are kept separately and only accessible to the investigators of this study or authorized researchers who consent to the End User License Agreement (EULA) governing usage of dataset. Some aspects of participants' electronic data were de-identified (e.g., faces blurred) as befit the permissions they gave.

\subsection{Participant recruitment}

We reached out to potential participants by posting flyers all over Purdue campus, through social networking services, and by word of mouth. The investigators also used official contact lists (such as of faculty, staff, and students in the department) to recruit participants through emails. When contacted by potential participants, investigators gave further information about the experiment, confirmed participants’ availability, and sent back a confirmation email. When participants arrived at the site of data collection, they were provided with the informed consent form, and investigators were responsible for answering all their questions and concerns. 

Thirty participants were recruited and participated in our data collection study during March and April 2022; however, only 21 had their data included in our dataset based on their permissions. All recruited participants were students and faculty/staff at Purdue University, and their ages range from 18 to 37 years old (mean = 24.3 years, S.D. = 5.2 years). In order to encourage participants' engagement in this experiment, we also compensated \$ 15 for each participation and also provided additional compensations (e.g., amazon gift cards as prizes) to three participants who had the highest scores in the given tasks.

\subsection{Stimuli design}

Video game scenarios such as N-back games and the Multi-Attribute Task Battery (MATB) are widely used for eliciting target cognitive workload levels but have limited relevance to real-world applications. In the interest of better mimicking real-world workload scenarios, we proposed a CCTV monitoring task scenario as the stimulus and collected physiological and behavioral signals under different workload levels. Such tasks, in which human operators monitor and control a display interface (or multiple interfaces simultaneously), are widely required in diverse human-machine system task scenarios such as security monitoring \cite{muller2008machine,wang2008affective}, air traffic management \cite{hoc1998cognitive,pacaux2002common, giraudet2015neuroergonomic} and performance checking \cite{liu2006queueing,stowers2017framework}. 

\begin{figure}[t]
    \centering
    \begin{subfigure}{0.9\linewidth}
        \centering 
        \includegraphics[width=1\linewidth]{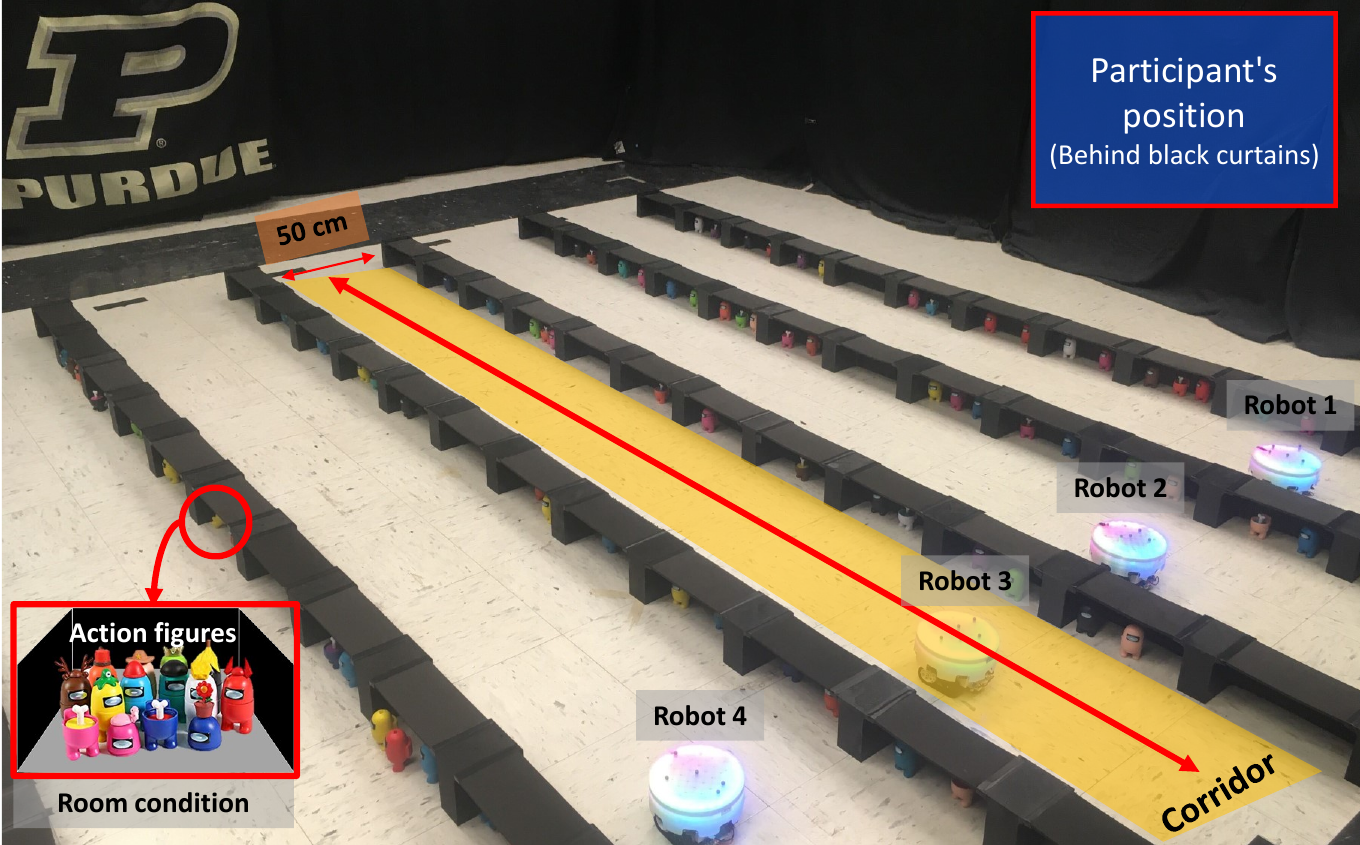}
        \caption{}
        \label{img:exp_environment}
    \end{subfigure}
    
    \begin{subfigure}[b]{0.365\linewidth}
        \centering 
        \includegraphics[width=1\linewidth]{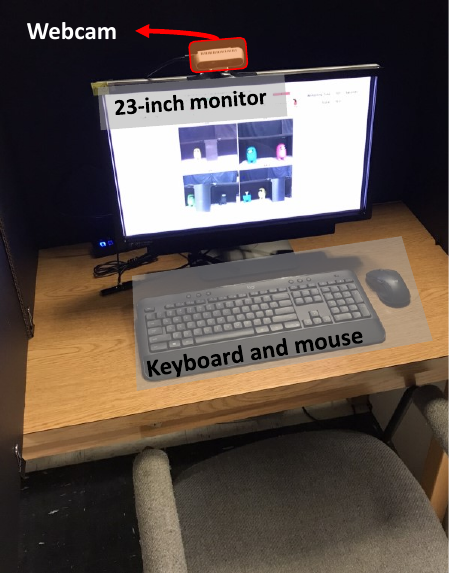}
        \caption{}
        \label{img:participant_environment}
    \end{subfigure}
    \begin{subfigure}[b]{0.515\linewidth}
        \centering 
        \includegraphics[width=1\linewidth]{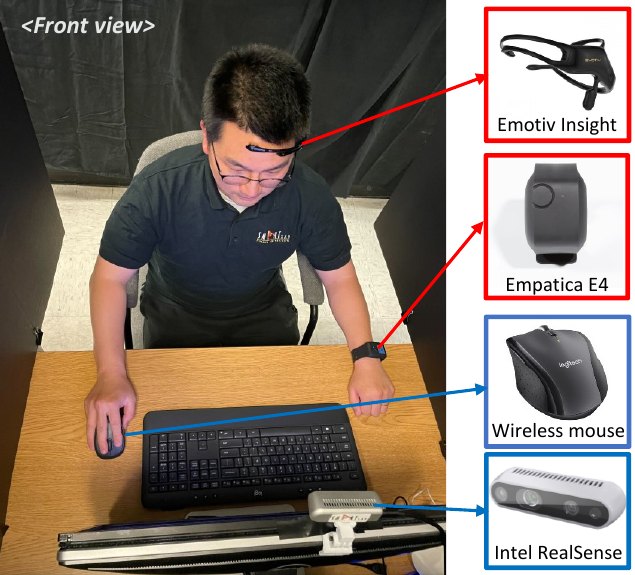}
        \caption{}
        \label{fig:device}
    \end{subfigure}
    \caption{Illustration of the designed CCTV monitoring task scenario, consisting of (a) four patrol robots capturing real-time monitoring video streams, (b) the desk at which participants conducted the surveillance tasks, and (c) the wearable sensors and devices used to collect physiological (red) and behavioral signals (blue) from each participant. }
    \label{fig:path}
\end{figure}

\begin{figure}[h]
    \centering
    \begin{subfigure}[b]{0.85\linewidth}
        \centering
        \includegraphics[width=1\linewidth]{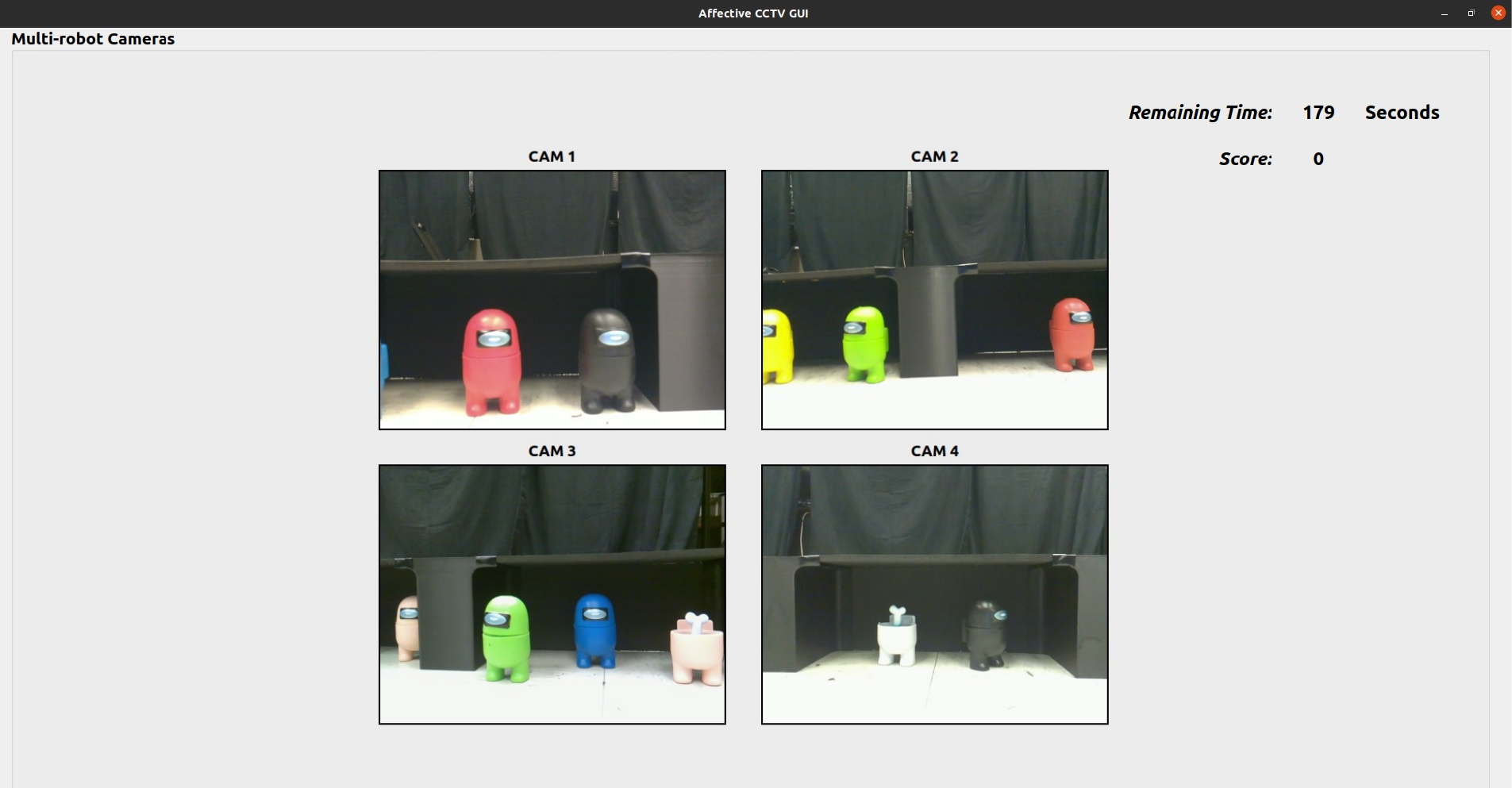}
        \caption{}
        \label{fig:GUI}
    \end{subfigure}
    
    \begin{subfigure}[b]{1\linewidth}
        \centering 
            \begin{subfigure}[b]{0.40\linewidth}
            \centering 
            \includegraphics[width=1\linewidth]{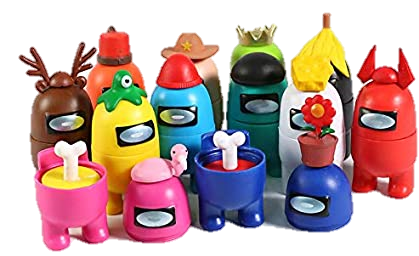}
            \caption{}
        \label{img:all_object}
        \end{subfigure}
        \begin{subfigure}[b]{0.15\linewidth}
            \centering
            \includegraphics[width=1\linewidth]{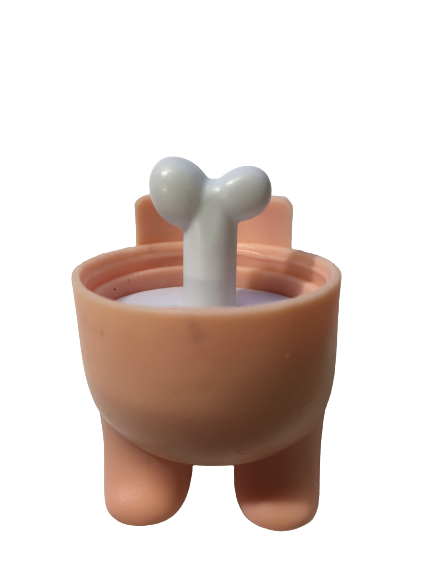}
            \caption{}
            \label{img:wrong_object}
        \end{subfigure}
        \begin{subfigure}[b]{0.16\linewidth}
            \centering 
            \includegraphics[width=1\linewidth]{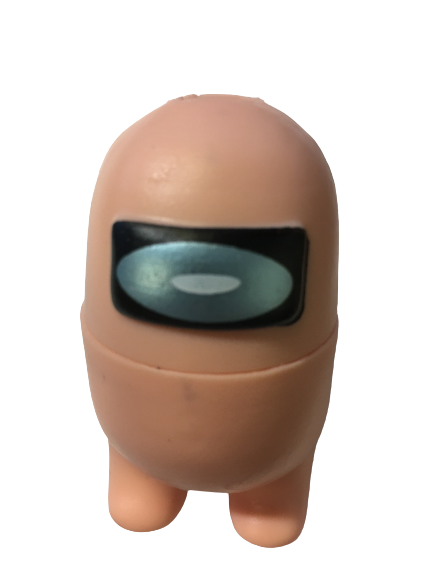}
            \caption{}
            \label{img:right_object}
        \end{subfigure}

    \end{subfigure}
    
    \caption{Illustration of designed CCTV monitoring task showing (a) the graphical user interface used by participants to conduct monitoring \cite{jo2024smart}, in which the camera views to be clicked on were placed in the center while the time remaining and obtained score were presented at the top right; and (b) representative objects to be monitored and recognized by participants, including (c) abnormal objects, and (d) normal objects.}
    \vspace{-13pt}
    \label{fig:object}
\end{figure}

Here, participants were asked to monitor multiple video streams captured by multiple patrol robot platforms passing by multiple separated rooms located in multiple corridors (Fig. \ref{fig:path}). As depicted in Fig. \ref{fig:object}, participants were asked to monitor and detect two types of objects: (a) abnormal objects and (b) normal objects. Each room contained a random number of objects with a random assortment of types. 
The number and speed of robots could be changed to present different levels of stimuli (e.g., low, medium, and high workloads). 
The thresholds can be determined based on the number and speed of robots. For instance, a low cognitive workload level can be achieved with a small number of robots moving at a slow pace, while a medium cognitive workload level can be achieved with either two robots or faster-moving robots. On the other hand, a high cognitive workload level can be achieved by using many fast-moving robots. These thresholds were determined through multiple beta and pilot tests, which took into account factors such as the number and speed of robots.

During the task, the participant was asked to observe the webcam streams captured by the robots and to use a mouse to click on camera views that contain any abnormal objects; the associated graphical user interface (GUI) is shown in Fig. \ref{fig:GUI} (for more information refer to \cite{jo2024smart}). 
 
To incentivize accurate identification and acknowledge participants' attentiveness and precision, we have designed a scoring system in which a participant is awarded 1 point for correctly identifying and clicking on an abnormal object. In contrast, failing to identify and click on an abnormal object results in a deduction of 3 points, which is a more severe penalty than the reward for correct identification. This scoring system aims to emphasize the importance of avoiding errors in the task and to motivate participants to be vigilant and precise in identifying abnormal objects by balancing rewards for correct actions and penalties for incorrect actions.


The designed CCTV monitoring task was conducted in a room, approximately 5 $m$ x 6 $m$ x 3 $m$ (width x depth x height), with a multi-robot system involving at most four mobile robot platforms \cite{jo2022smartmbot} performing the patrol task (see Fig. \ref{img:exp_environment}) to provide video streams; also in the room was a desk supporting a 23-inch monitor, a common wireless keyboard, and a mouse (see Fig. \ref{img:participant_environment}) with which participants performed the monitoring task. During each collection period, only one participant was in the room. participants were not allowed to directly observe the multi-robot system during the experiment, but could hear the sounds generated by its movements. However, we believe that the impact of these sounds on participants' performance is minimal, as the sounds generated were consistent and unobtrusive, avoiding significant distractions. Moreover, the participants were informed about the presence of these sounds before the experiment, allowing them to acclimate to the environment. Furthermore, we intentionally included the presence of noise to simulate real-world conditions, thereby increasing the ecological validity of our study.


\subsection{Measures}
We assessed participants' subjective cognitive workload using both ISA \cite{jordan1992instantaneous} and NASA-TLX \cite{hart1988development} with weights, and also measured their subjective emotional state using SAM \cite{bradley1994measuring}.
The ISA is a quick and simple assessment tool to measure mental workload with only five items (e.g., Underutilized, Relaxed, Comfortable, High, and Excessive). The NASA-TLX is a widely used assessment tool in physiology fields to measure task loads from five dimensions with 7-point scales on mental/physical/temporal demand, performance, effort, and frustration. The SAM is a non-verbal emotion assessment tool that directly measures the three domains related to emotional states. The details of aforementioned three measures can be found in Table \ref{tab:annotation_for_survey}. 


\begin{table}[h]
    \centering
    \caption{Emotion, Cognitive load, and personality trait annotations.}
    \label{tab:annotation_for_survey}
    \resizebox{\linewidth}{!}{%
    \begin{tabular}{cccc}
        \hline         \hline
        \textbf{Categories} & \textbf{Name} & \textbf{Scale range} & \textbf{Description} \\ \hline
        Emotion & SAM & from -4 to 4 & 
        \begin{tabular}[c]{@{}l@{}}Two dimensions to assess emotional state: \\ valence (from negative to positive) \\ and arousal (from calm to excited).
        \end{tabular} \\ \hline
        Workload & ISA &
          \begin{tabular}[c]{@{}c@{}}from -2 to 2 \end{tabular} &
          \begin{tabular}[c]{@{}l@{}}Simple subjective workload assessment \\using five dimensions:\\ Underutilized (-2), Relaxed (-1), \\Comfortable (0) High (1), and Excessive (2).\end{tabular} \\ \hline
        Workload & NASA-TLX &
          \begin{tabular}[c]{@{}c@{}}from 1 to 7 \end{tabular} &
          \begin{tabular}[c]{@{}l@{}}Seven categories for measuring workloads; \\ Mental Demand, Physical Demand, \\ Temporal Demand,  Performance, Effort, \\ and Frustration. \end{tabular} \\ \hline
    \end{tabular}%
    }
\end{table}

\begin{figure*}[t]
    \centering
    \includegraphics[width=0.85\linewidth]{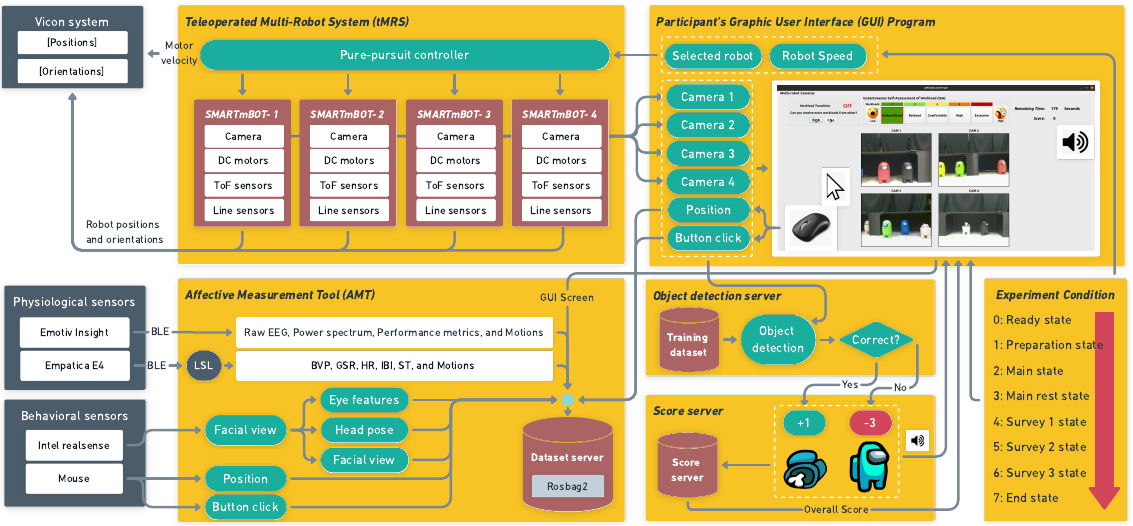}
    \caption{Overall system configurations used in this user experiment for data collection and storing process.}
    \label{fig:process}
\end{figure*}

\subsection{Apparatus of data collection}

During the task, participants were required to wear two off-the-shelf wearable biosensors for physiological data collection, and their behavioral data was recorded through a webcam (i.e., Intel RealSense D435i) mounted on the monitor and a mouse as shown in Fig. \ref{fig:device}:

\begin{itemize}
    \item Emotiv Insight – captured 5-channel electroencephalogram (EEG), power spectrum (POW) (i.e., theta, alpha, beta, and gamma), and performance metrics generated from the EmotivPro SDK.
    \item Empatica E4 Wristband – captured photoplethysmography (PPG), 9-axis acceleration, skin temperature (SKT), electrodermal activity (GSR, or EDA), heart rate (HR) and the inter-beat interval (IBI) derived from PPG.
    \item Intel Realsense D435i – captured participant's facial views and eye movements.
    \item Mouse – recorded participant's mouse movements. 
\end{itemize}

Additionally, participant's monitor screen, including CCTV video stream, mouse positions and status of pushed mouse buttons, was recorded for the experiment, enabling researchers to easily understand the status of the task performed by each participant. 
Table \ref{tab:data} summaries collected data in our dataset.

As illustrated in Fig. \ref{fig:process}, all sensors, devices and GUI programs used in the experiment were connected through Robot Operating System 2 (ROS2), where signals were collected as ROSbag2 files with synchronized time (e.g., ROS2 timestamp). The ROSbag2 format has more benefits than a traditional CSV format file in terms of collecting and analyzing the dataset \cite{jo2020rosbag}, since it can ensure to synchronize the recording all topic data and to easily analyze the dataset by replaying both using a single ROSbag2 file.

\subsection{Data collection procedures}
\begin{figure*}[t]
    \centering
    \includegraphics[width=0.85\linewidth]{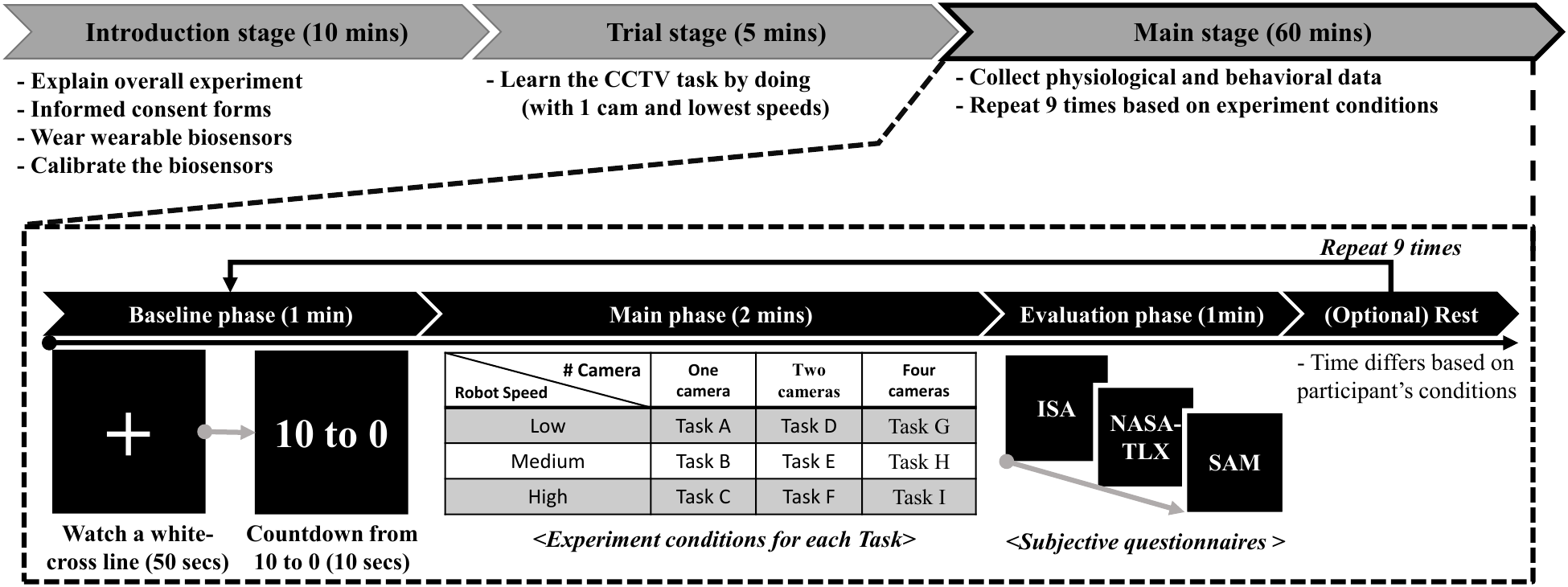}
    \caption{Overall procedures for the data collection in the experiment. The supplementary video about this procedure is able to be found at \url{https://youtu.be/BxVVj7R9b70}.}
    \label{fig:overall_procedures}
\end{figure*}

Each participant's data collection procedure was conducted in following three main stages: (1) Introduction stage, (2) Trial stage, and (3) Main stage. Fig. \ref{fig:overall_procedures} illustrates the overall data collection procedures. Additionally, there is a supplementary video at \url{https://youtu.be/BxVVj7R9b70} that explains the details of this experiment design and procedures. 

\begin{figure*}[t]
    \centering
    \begin{subfigure}[b]{0.33\linewidth}
        \includegraphics[width=1\linewidth]{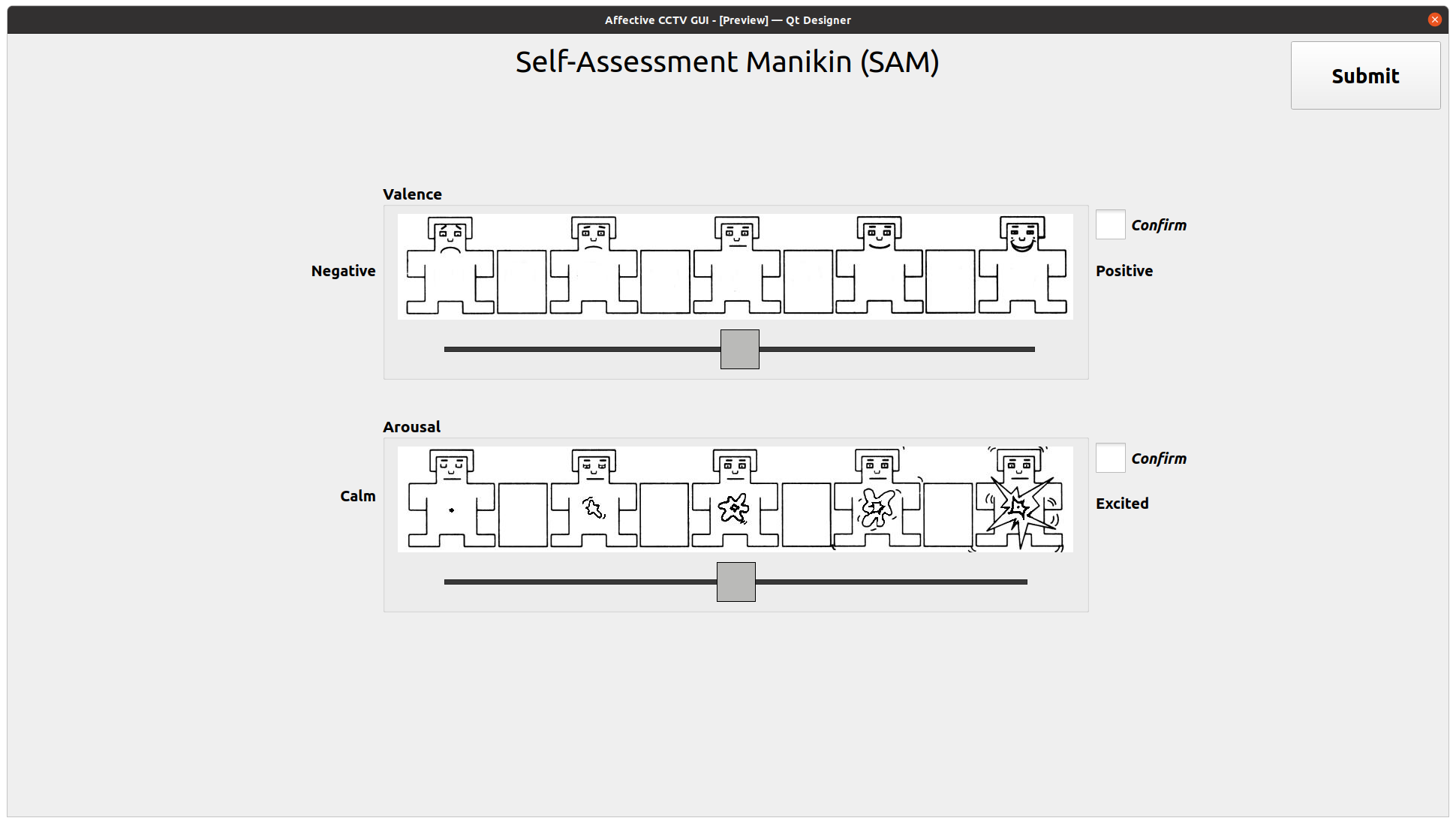}
        \caption{SAM questionnaire}
        \label{img:sam_gui}
    \end{subfigure}
    \begin{subfigure}[b]{0.33\linewidth}
        \centering 
        \includegraphics[width=1\linewidth]{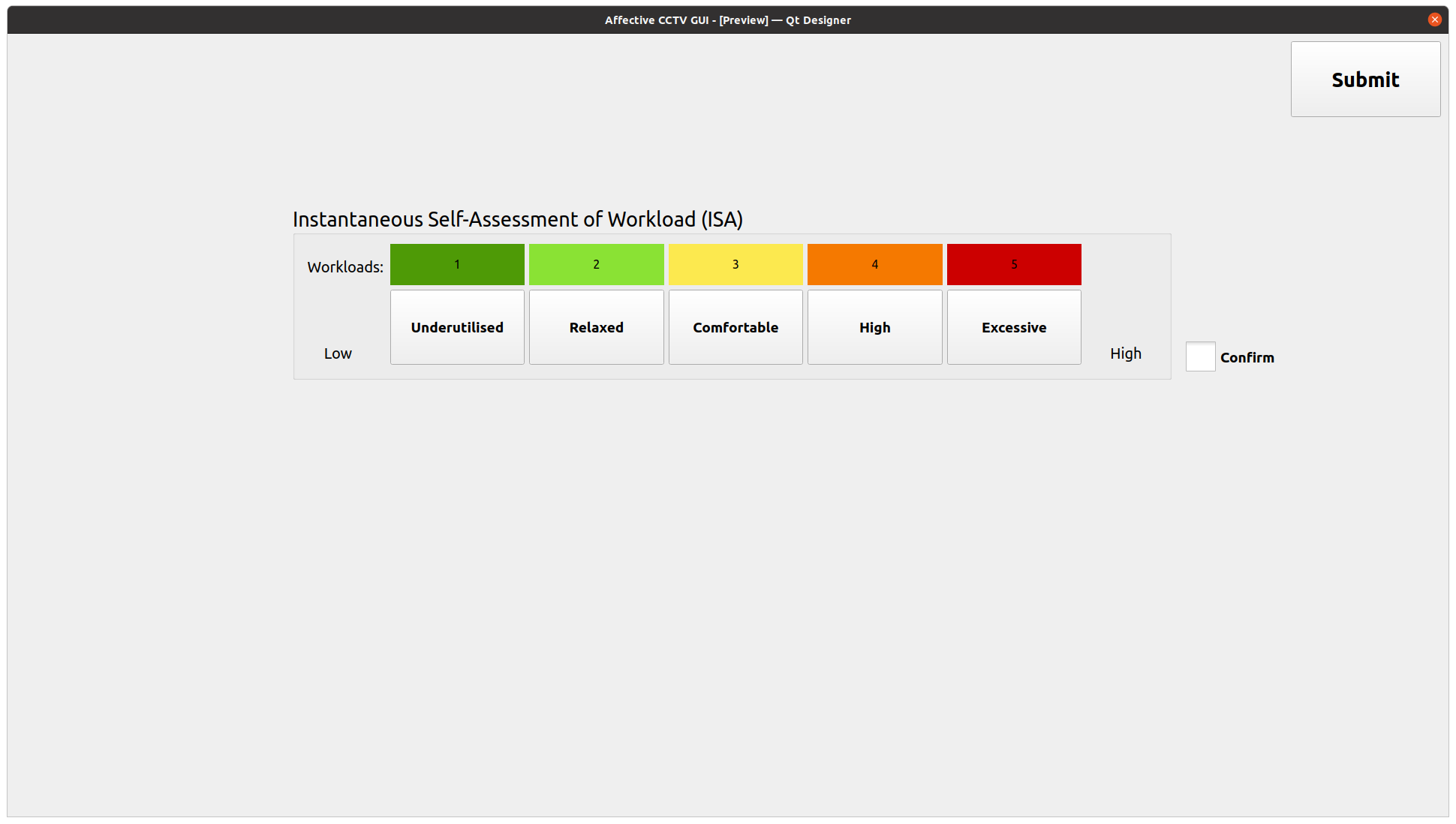}
        \caption{ISA questionnaire}
        \label{img:ISA_gui}
    \end{subfigure}
    \begin{subfigure}[b]{0.33\linewidth}
        \centering 
        \includegraphics[width=1\linewidth]{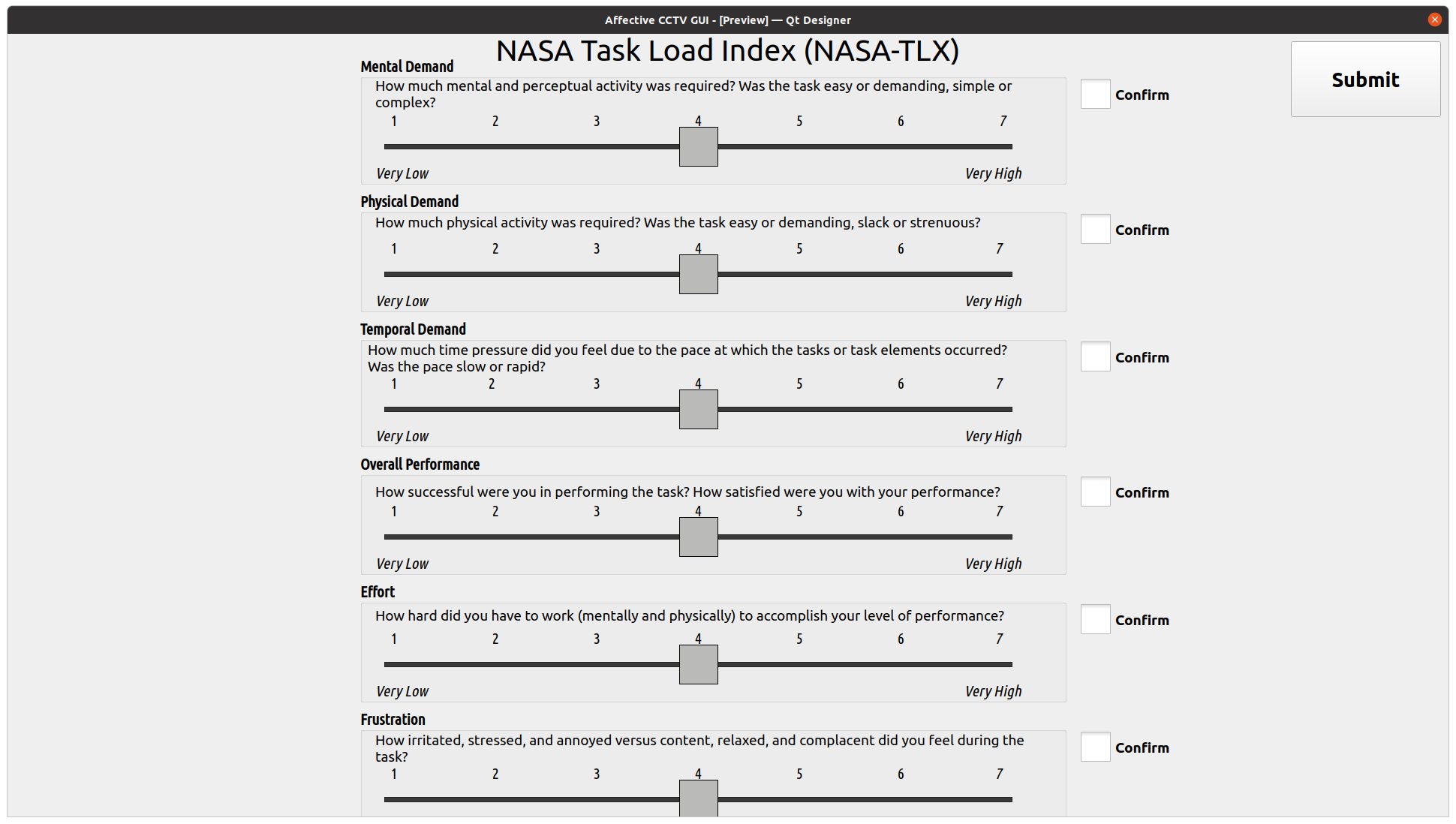}
        \caption{NASA-TLX questionnaire}
        \label{img:nasa_tlx_gui}
    \end{subfigure}
    \caption{Graphical user interfaces for subjective questionnaires used in the user experiment: (a) SAM for measuring emotional states using Arousal and Valence, (b) ISA, and (c) NASA-TLX for measuring workloads.}
    \label{fig:subjective_questionaires}
\end{figure*}

\textbf{Introduction stage:} At the beginning of the data collection, investigators checked again if the participant satisfied the qualifications for this study (such as not having medical history on mental, heart disorders, skin allergies, etc.). The satisfied participants then read and signed the informed consent, and filled out demographic and personality questionnaires. The demographic questionnaire collected participants' age, gender, level of education, medical history, experiences in conducting video-based surveillance or monitoring tasks, and daily usages of interacting with a monitor that indicated their capability and ability. The personality questionnaire was based on Big Five Personality Test to categorize participants' personal traits into five personality traits using IPIP Big-Five Factor Markers \cite{goldberg1990alternative} as follows:

\begin{itemize}
    \item Extraversion: a participant who has high scores tends to be outgoing/talkative/social, whereas one who has low scores tends to be reflective and reserved behavior. 
    \item Agreeableness: a participant who has high scores tends to be friendly and optimistic, whereas one who has low scores tends to be critical and aggressive. 
    \item Conscientiousness: a participant who has high scores tends to be careful and diligent, whereas one who has low scores tends to be impulsive and disorganized.  
    \item Emotional stability (or neuroticism): a participant who has high scores tends to be sensitive and nervous, whereas one who has low scores tends to be resilient and confident. 
    \item  Intellect/Imagination (or openness to experience): a participant who has high scores tends to be inventive and curious, whereas one who has low scores tends to be traditional and conventional. 
\end{itemize}

After that, investigators outlined the experimental process and gave instructions for each of the tasks that the participants need to complete, and then helped each participant wear and calibrate the physiological and behavioral sensors.

\textbf{Trial stage:} Following the introduction stage, participants were given time to get familiar with the hardware and software utilized and understand the tasks of this experiment. They conducted a trail experiment with one camera view and the minimum speed of the robots, which would not be included in the main experiment again. 
This stage was continued until participants fully understood and got familiar with the CCTV monitoring task in this study.

\begin{figure*}[t] 
    \centering
    \begin{subfigure}[b]{0.33\linewidth}
        \includegraphics[width=1\linewidth]{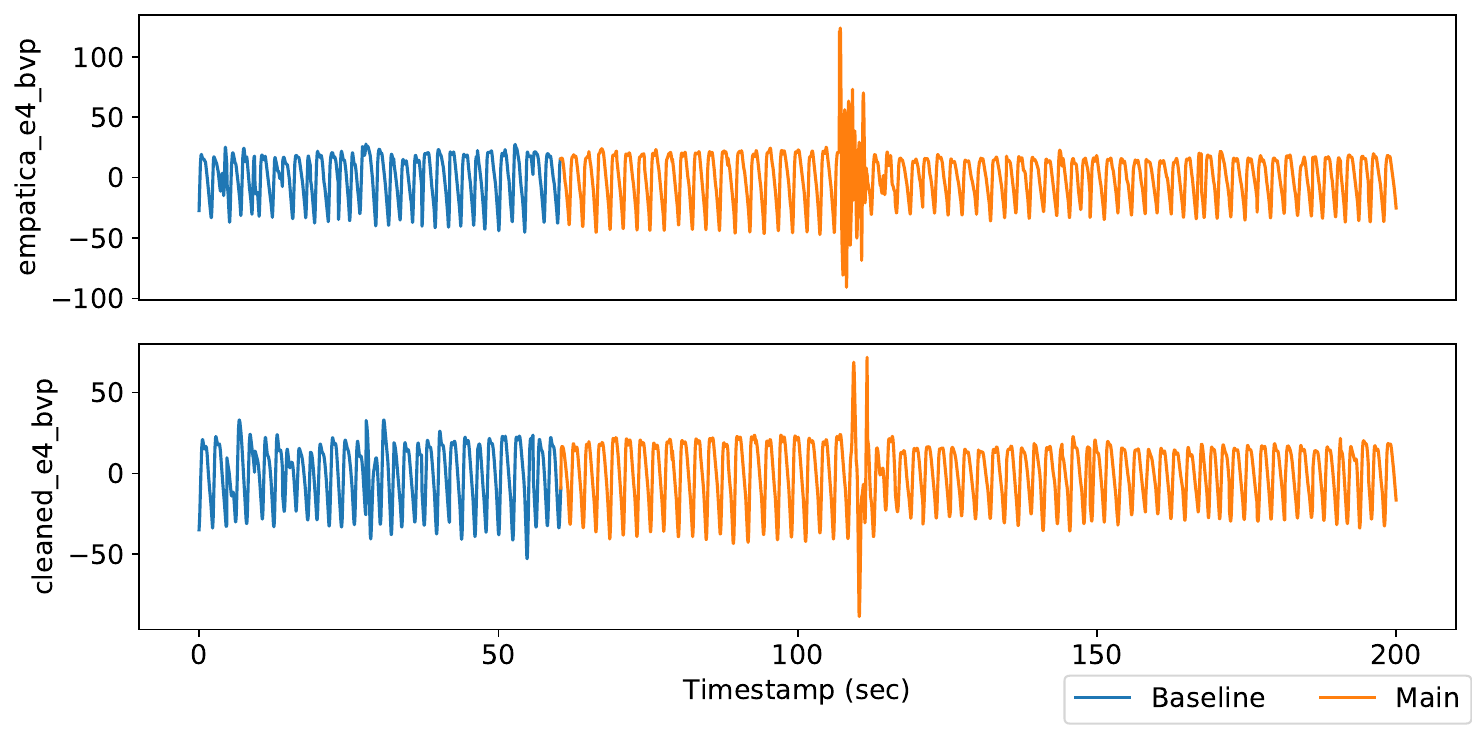}
        \caption{Raw (top) and cleaned BVP data (bottom) of Empatica E4}
        \label{img:raw_clean_bvp}
    \end{subfigure}
    \begin{subfigure}[b]{0.33\linewidth}
        \centering 
        \includegraphics[width=1\linewidth]{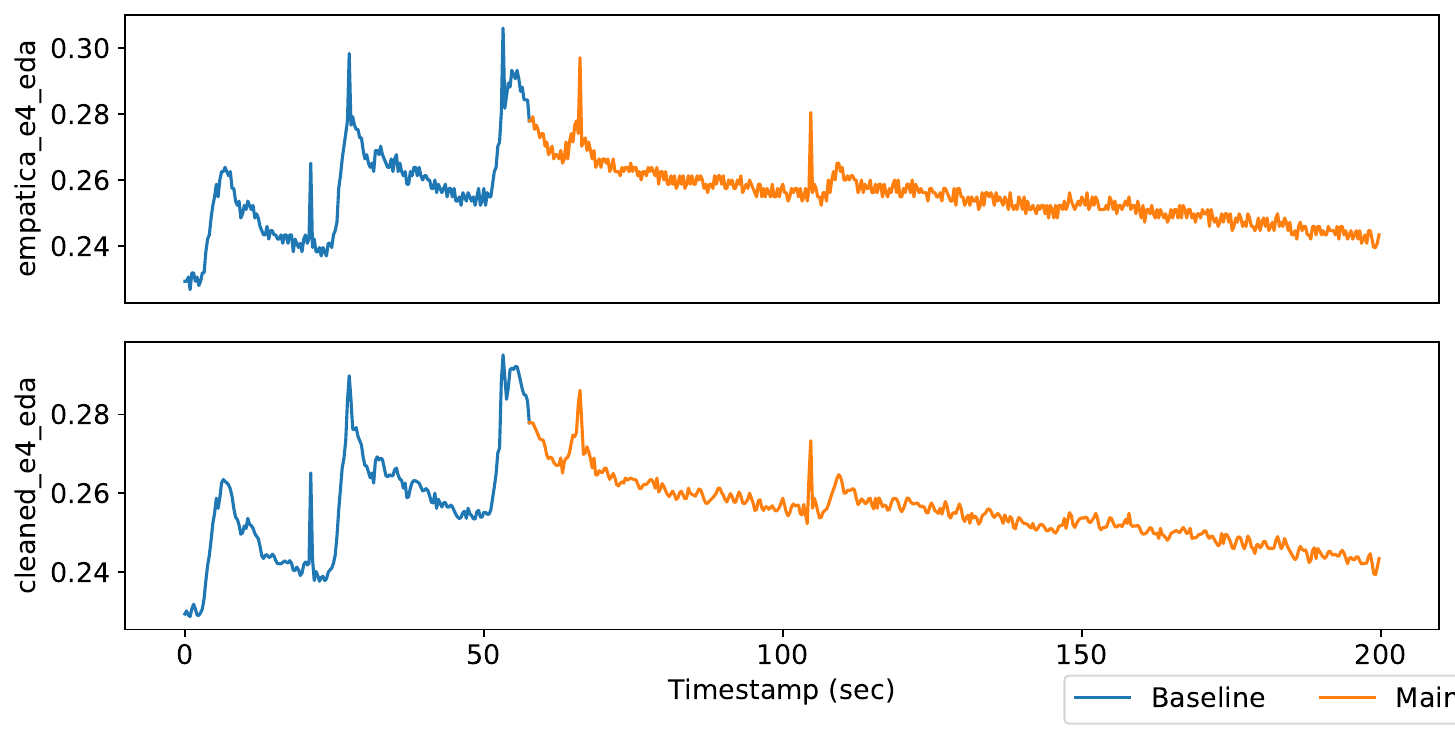}
        \caption{Raw (top) and cleaned GSR data (bottom) of Empatica E4}
        \label{img:raw_clean_eda}
    \end{subfigure}
    \begin{subfigure}[b]{0.33\linewidth}
        \centering 
        \includegraphics[width=1\linewidth]{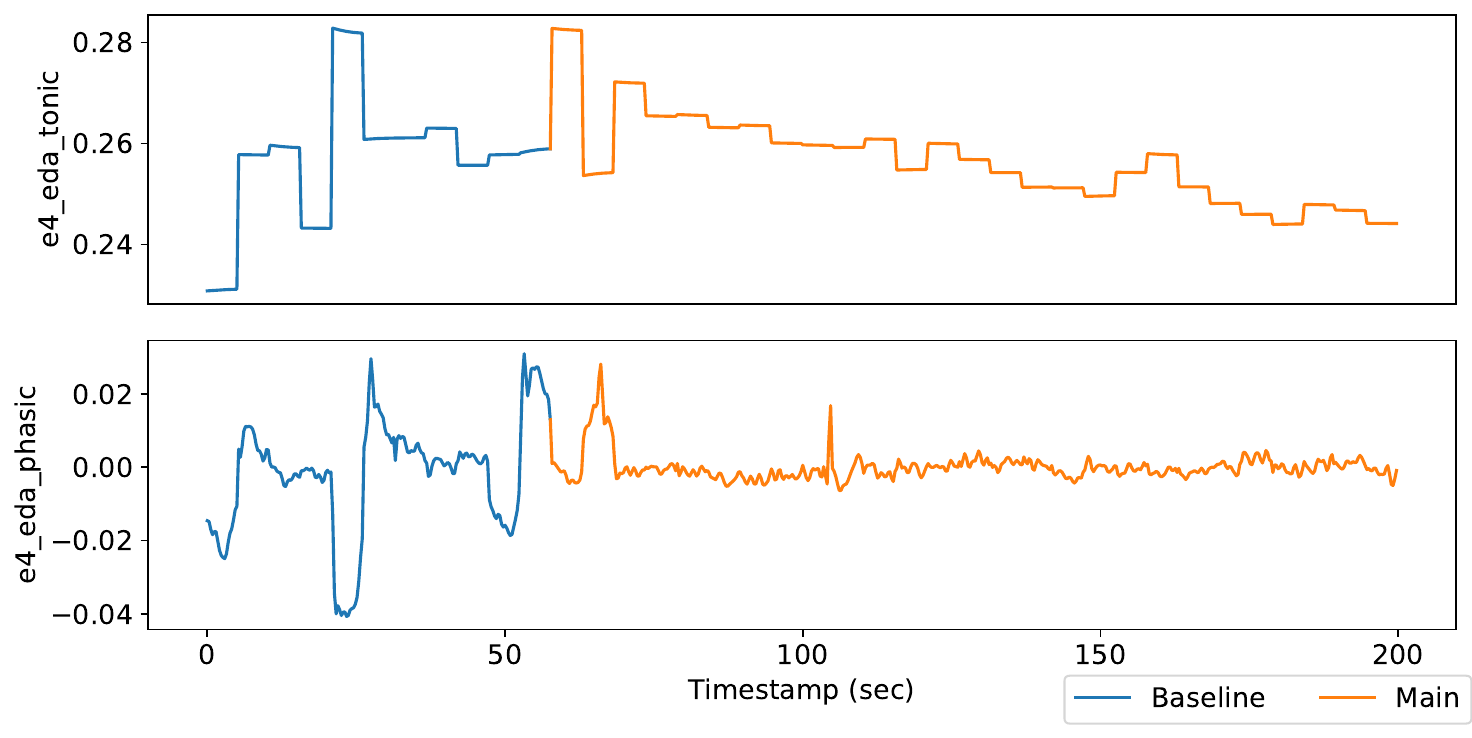}
        \caption{Tonic (top) and Phasic features (bottom) of the GSR data}
        \label{img:eda_features}
    \end{subfigure}
    
    \begin{subfigure}[b]{0.42\linewidth}
        \centering 
        \includegraphics[width=1\linewidth]{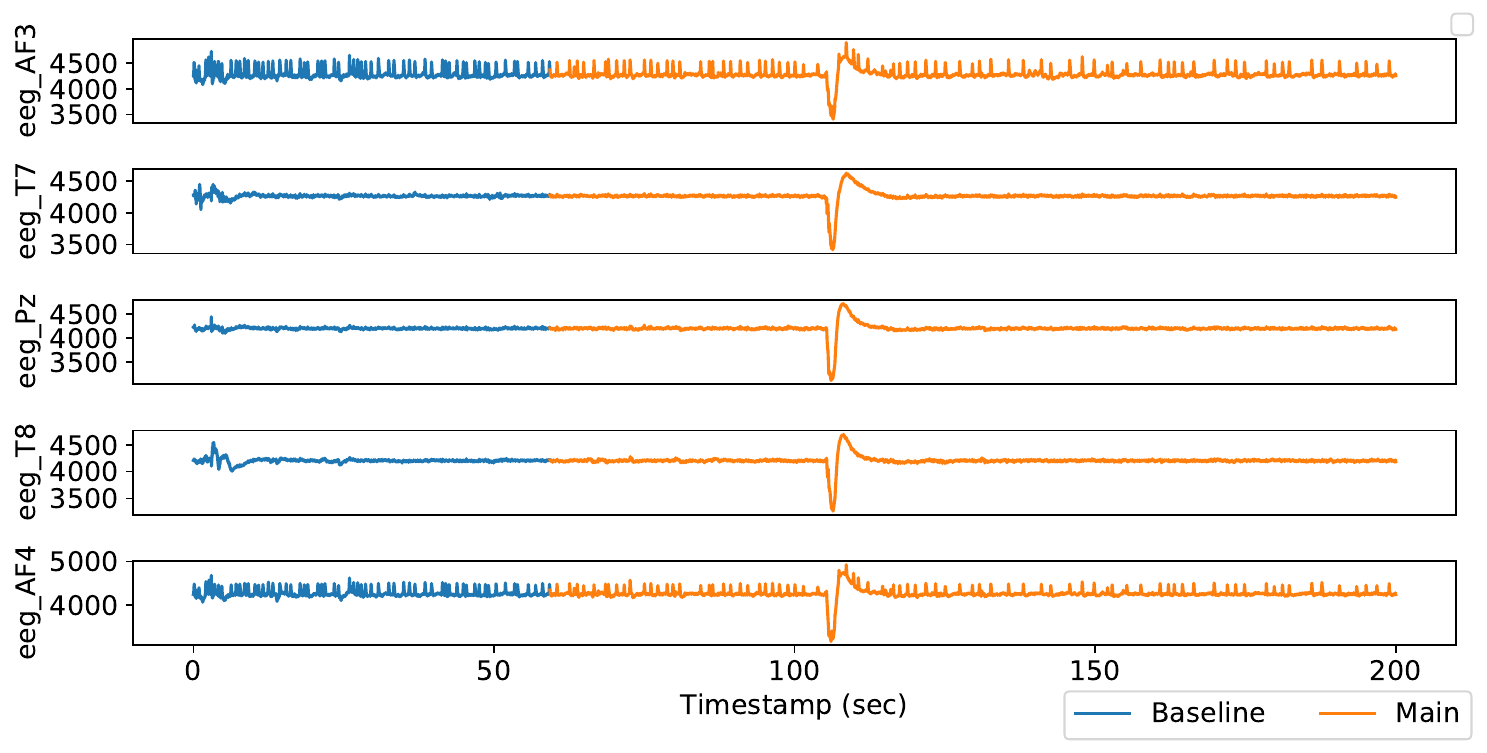}
        \caption{Raw 5-channels EEG data of Emotiv Insight}
        \label{img:raw_eeg}
    \end{subfigure}
    \begin{subfigure}[b]{0.285\linewidth}
        \centering 
        \includegraphics[width=1\linewidth]{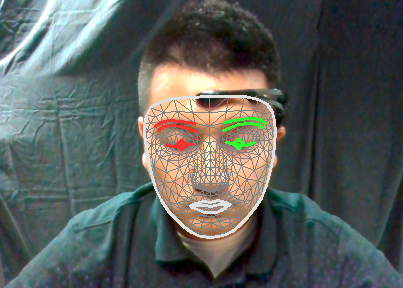}
        \caption{Facial features from closed eyes}
        \label{img:facial_expression_1}
    \end{subfigure}
    \begin{subfigure}[b]{0.27\linewidth}
        \centering 
        \includegraphics[width=1\linewidth]{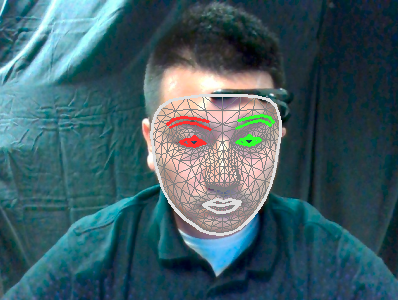}
        \caption{Facial features from open eyes}
        \label{img:facial_expression_2}
    \end{subfigure}
    \caption{Examples of the raw and cleaned collected physiological and behavioral data: raw and cleaned (a) BVP data and (b) GSR data,
    (c) features extracted from the GSR data (tonic and phasic components), (d) raw EEG signals, and behavioral data extracted from the facial videos with (e) closed eyes and (f) open eyes.} The blue lines in (a) - (d) mean the collected data in the baseline phase and the orange lines mean the collected data in the main phase.
    \label{fig:example_data}

\end{figure*}
\textbf{Main stage:} The main stage has four repeated sub-phases as illustrated in the dashed box in Fig. \ref{fig:overall_procedures}: baseline phase (about 1 minute), main phase (about 2 minutes), evaluation phase (about 1 minute), and rest phase (optional). 
In each baseline phase, participants watched a white cross-line with a black background for 50 seconds, and then a countdown from 10 to 0, which helped participants minimize cognitive workload affected by their previous conditions \cite{van2003mental, hjortskov2004effect, liu2021taking, lee2021investigating}. 

In each main phase, the participant operated one CCTV monitoring task selected from nine tasks at different workload levels decided by different combinations of three different numbers of camera views (one, two, or four cameras) and three different speeds of the multi-robot system (low, medium, and high speed).

During each rest phase, participants were given a break if they requested one due to pain caused by the wearable sensors or cognitive load from the tasks. At that time, we provided enough time for participants to relieve stress and pain by removing the wearable sensors. Other participants who did not need rest time continued with the remaining tasks. Since this phase was optional, we did not collect physiological and behavioral signals during this period. Additionally, before conducting the next set of tasks, we changed experimental environment by randomly selecting new positions for the robots and running new experiment programs to collect data. This process took a minimum of 1 minute.
The order of nine tasks was randomly selected for each participant. After completing the assigned monitoring task and before the rest phase, the participant moved to the evaluation phase to use the GUI-based questionnaires as shown in Fig. \ref{fig:subjective_questionaires} to report subjective cognitive workload via ISA and NASA-TLX respectively, and subjective emotion state via SAM. After finishing the above three sub-phases, the participant could ask for a rest based on his/her conditions. When the participant decided to continue, the experiment would start to repeat these four sub-phases.

\section{Data Records}
\label{sec:dataset_sum}
\subsection{Dataset summary}
Table \ref{tab:summary_dataset} summarizes the MOCAS, which contains multimodal data from 21 participants, including physiological signals, facial camera videos, mouse movement, screen record videos, and subjective questionnaires. Fig. \ref{fig:example_data} shows examples of physiological and behavioral data from the raw dataset. The total size of the dataset is about 722.4 GB which includes 754 ROSbag2 files. 
\begin{table*}[!t]
    \centering
    \caption{Summary of the MOCAS dataset contents.}
    \label{tab:summary_dataset}
    \resizebox{1\linewidth}{!}{%
    \begin{tabular}{c|l}
    \hline\hline
    \textbf{Number of participants} &
      21 (7 females and 14 males) \\ \hline
    \textbf{Age} &
      from 18 to 37 (mean=24.3 age, std.=5.2 age) \\ \hline
     \begin{tabular}[c]{@{}c@{}}\textbf{Average time} \\ \textbf{per experimental task} \end{tabular}     
    &
     mean= 3.52 mins, std.=0.28 mins \\ \hline
     \textbf{Personality traits} &
      Extraversion, Emotional stability, Agreeableness, Conscientiousness, and Intellect/Imagination \\ \hline
    \textbf{Physiological signals} &
      \begin{tabular}[c]{@{}l@{}}Empatica E4: BVP, GSR, HR, IBI, SKT\\ Emotiv Insight: raw 5-channel EEGs, EEG band powers (i.e., $\alpha, \beta, \gamma,$ and $\theta$), performance metrics 
      \end{tabular} \\ \hline
    \textbf{Behavioral features} &
      \begin{tabular}[c]{@{}l@{}}facial view (30 Hz), facial features \& expressions (30 Hz), \\ and Mouse positions \& button clicking status (True or False)\end{tabular}
      \\ \hline
    \textbf{Experiment status} &
      \begin{tabular}[c]{@{}l@{}}0=loading phase, 1=baseline phase, 2=main phase, 4=Evaluation phase (SAM), \\ 5=Evaluation phase (ISA), 6=Evaluation phase (NASA-TLX), and 7=Score phase\end{tabular} \\ \hline
      \textbf{Experiment recording} &
      \begin{tabular}[c]{@{}l@{}}Screen record video (30 Hz)\end{tabular} \\ \hline
          \textbf{Experiment scores} &
      Obtained scores, Success click, Failure click, and Success rate (=success clicks/all clicks) \\ \hline
    \textbf{Subjective annotations} &
      \begin{tabular}[c]{@{}l@{}}SAM: two categories; valence (from negative to positive) and \\ arousal domain (from calm to excited) with a range from -4 to +4\\ ISA: five categories; Underutilized (-2), Relaxed (-1), Comfortable (0), High (1), \\ and Excessive (2)\\ NASA-TLX: Seven categories for measuring workloads; Mental demand,\\ Physical demand, Temporal demand, Performance, Effort, and Frustration \\ with a range from 1 to 7\end{tabular} \\ \hline

    \end{tabular}%
    }
\end{table*}
\begin{figure}[t]
    \centering
    \includegraphics[width=1\linewidth]{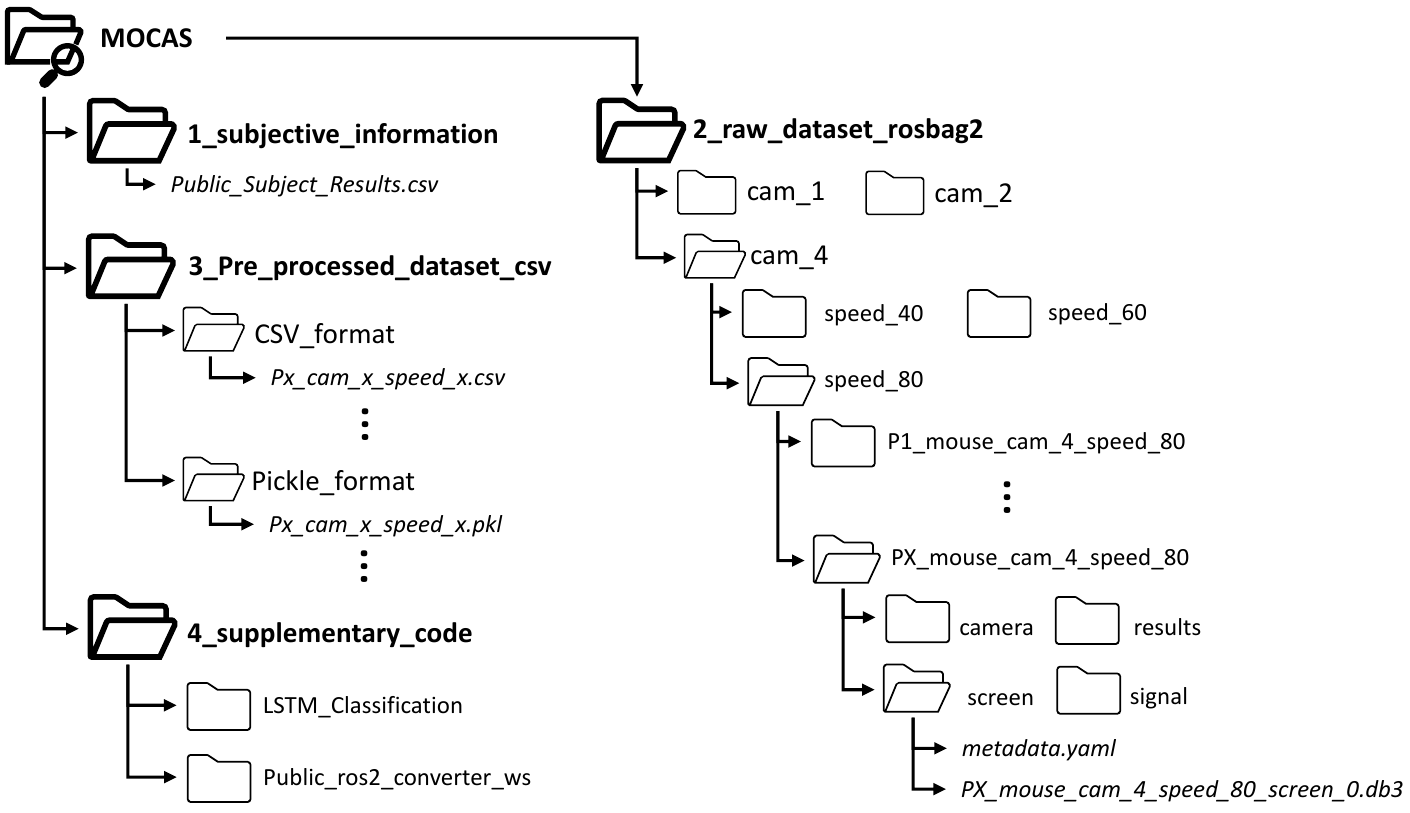}
    \caption{A folder tree of the online repository that displays directory paths and files for the MOCAS dataset.}
    \label{fig:folder_tree}
\end{figure}

Fig. \ref{fig:folder_tree} shows a folder tree of the online repository that displays directory paths and files for the MOCAS dataset. There are three major folders based on the types of the dataset:
\begin{itemize}
    \item \textit{\textbf{1\_subjective\_information}}: a single csv file containing subject's demographic (e.g., age, gender, personality trait, and experience), answers of the subjective questionnaires used in this experience, and performance (e.g., scores, success click, failure click, and success rate). 
    
    \item \textit{\textbf{2\_raw\_dataset\_rosbag2}}: raw rosbag2 files saved into each sub-folder depending on the experimental factors (e.g., robot speed and the number of camera views) and data types (e.g., camera, signal, results, and screen). The last sub-folder of this major folder has two files: \textit{$metadata.yaml$} and \textit{ $*.db3$}.
    
    \item \textit{\textbf{3\_pre-processed\_dataset}}: downsampled CSV-format and Pickle-format \cite{van_rossum_2020,fasnacht2018mmappickle} files with 100 Hz sampling rate. The overall size of the CSV files and Pickle file is 51.9 GB and 32.8 GB, respectively. Each file has the annotations of the subjective questionnaires, the number of camera views, robot speeds, features of the physiological and behavioral data, and filtered physiological signals.

    \item \textit{\textbf{4\_supplementary\_code}}: contains the converter codes from raw ROSbag2-format into pickle and csv-format files, and supplementary codes used on this paper. The codes are written by Python under Ubuntu 20.04.
\end{itemize}

\subsection{Pre-processed and downsampled data} \label{sec:preprocessed_down}
The raw ROSbag2 format files were converted to Pickel and CSV format with 100 Hz sampling rate, which were saved to two different folders based on the type of file format (i.e., CSV and Pickle \cite{van_rossum_2020}). Each file contains raw signals, each channel's EEG signal divided from chunk EEG signals, cleaned BVP and GSR signals, and physiological features from the GSR signals (e.g., tonic and phasic domains).

For cleaning the raw BVP and GSR signals, we applied a bandpass filter to remove the noise from both the raw BVP and GSR signals through \cite{Makowski2021neurokit}. 
For the physiological features, we extracted tonic and phasic features from the raw GSR signals. They can provide valuable insights into the nervous system activity and emotional arousal of an individual \cite{wang2018arousal}. The tonic feature is the baseline level of a physiological signal called Phasic Skin Conductance Response (SCR). To extract the tonic component from the raw GSR signal, we applied a moving averages algorithm for baseline correction to refine the tonic component. The phasic feature refers to the rapid and temporary changes in physiological signals that occur in response to specific stimuli or events, called phasic skin conductance responses (SCRs), which often associated with brief fluctuations in arousal or attention. To identify phasic events, we use peak detection algorithms that detect peaks in the GSR signal. The prominent peaks usually represent the SCRs by analyzing amplitude, rise time, and half-recovery time.

Furthermore, each file contains behavioral features from facial videos (e.g., Action units (AUs), probability and types of facial expression \cite{boyko2018performance}, and Eye Aspect Ratio (EAR)), as well as the results of the self-reporting questionnaires.

For extracting AUs and EAR, we first applied the facial detection algorithm to extract the face area from the facial videos and extracted each facial feature from the face area, called AUs (see Fig \ref{img:facial_expression_1} and Fig \ref{img:facial_expression_2}), through a face landmark ($L$) detection of Google MediaPipe \cite{lugaresi2019mediapipe}. Then, the EAR of each eye is calculated using six landmarks around the eye. The average EAR of eyes is measured using the following Eq. \ref{eq:ear_eq} \cite{thiha2023efficient}:

\begin{equation}
    \label{eq:ear_eq}
    \begin{gathered}
EAR_{left} = \frac{\|L_{160} - L_{144}\| + \|L_{158} - L_{153}\|}{2\| L_{33} - L_{133} \|}\\  
EAR_{right} = \frac{\|L_{385} - L_{380}\| + \|L_{387} - L_{373}\|}{2\| L_{362} - L_{263} \|}\\
EAR = (EAR_{left} + EAR_{right})/2
    \end{gathered}
\end{equation}

The details of all data in the pre-processed MOCAS dataset can be found in Table \ref{tab:data}.

\begin{table*}[!t]
    \centering
    \caption{Details of the data in the pre-processed MOCAS dataset.}
    \label{tab:data}
    \resizebox{1  \linewidth}{!}{%
    \begin{tabular}{c|lccl}
         \hline\hline\textbf{ Device/Source} & \textbf{Collected data} & \textbf{Channels} & \textbf{Sampling rate } & \textbf{Signal range [min, max]} \\
        \hline Empatica E4 Wristband & BVP (PPG) & 1 & $64 \mathrm{~Hz}$ & N/A \\
        \cline { 2 - 5 } & GSR & 1 & ~~~$4 \mathrm{~Hz}$ & $[0.01 \mu \mathrm{S}, 100 \mu \mathrm{S}]$ \\
        \cline { 2 - 5 } & HR (from BVP) & 1  & $1 \mathrm{~Hz}$ & N/A 
        \\
        \cline { 2 - 5 } & IBI (from BVP) & 1 & $1 \mathrm{~Hz}$ & N/A 
        \\
        
        \hline Emotive Insight & EEG (with Contact Quality) & 6 & $128 \mathrm{~Hz}$ & N/A \\
        \cline { 2 - 5 } & EEG band powers & 25 & ~$ 8 \mathrm{~Hz}$ & {$[0, 100]$} \\
        \cline { 2 - 5 } & Performance metrics &   7  & ~~~$1 \mathrm{~Hz}$ & {$[0, 1]$} \\
        \hline
        
        \hline Intel RealSense D435i & Action Unit (AU) & 3 (x, y, average) & $1 \mathrm{~Hz}$& N/A\\
        \cline{2 - 5 }  & Eye Aspect Ratio (EAR) & 3 (left, right, average)& $1 \mathrm{~Hz}$& N/A\\
        
        \hline Subjective Annotations &  Cognitive Load (from ISA) & 1 & N/A & [-2,2]  \\
        \cline{2 - 5 } &  Cognitive Load (from NASA-TLX-Weighted) & 1 & N/A & [0,100]\\
        \cline{2 - 5 } &  Emotion (from SAM) & 2 (arousal, valence) & N/A & [-4,4]\\
        \hline GUI program & Experiment states &  1 & N/A & $[0, 7]$ \\
        
        \hline
    \end{tabular}
    }
\end{table*}

\section{Analysis of dataset}
\subsection{Correlation between personality traits and physiological and behavioral signals}

\begin{figure}[t]
    \centering
    \includegraphics[width=1\linewidth]{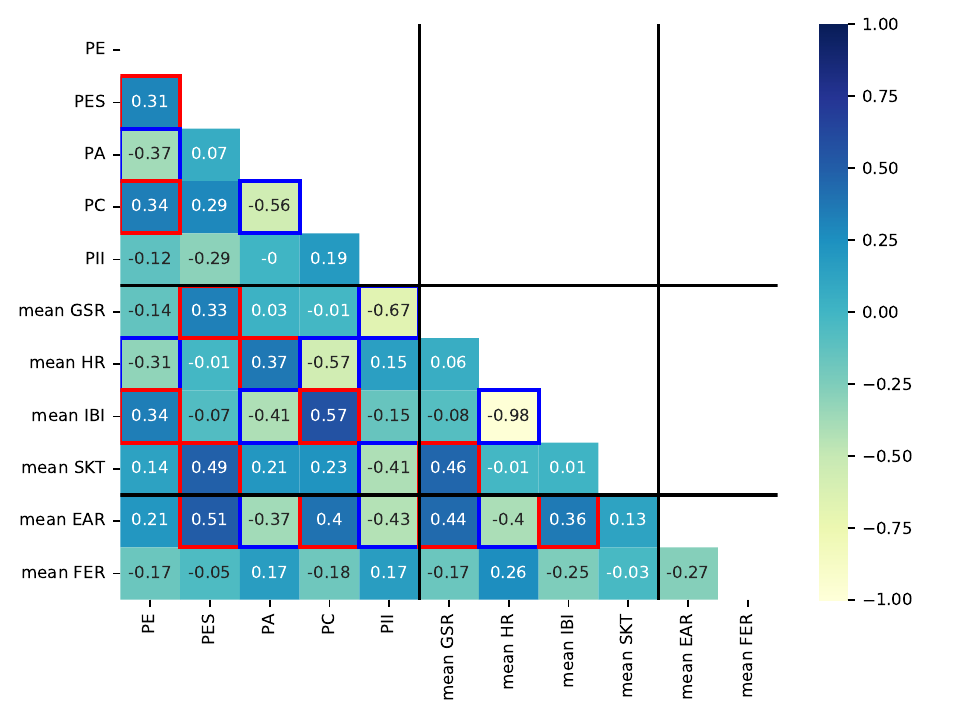}
    \caption{Correlation matrix for the correlation between personality traits and physiological and behavioral features. The red box means the moderate positive correlation between the two variables, whereas the blue box means the moderate negative correlation.}
    \label{fig:correlation_img}
\end{figure}

We calculated Pearson's correlation coefficient ($\gamma$) to find the relationship between the participant's personal traits and the mean of their physiological and behavioral data. $\gamma$ indicates the relationship between two variables with ranges from -1 to 1. The positive high $\gamma$ means a proportional relationship between both variables, whereas the negative high $\gamma$ means an inverse relationship \cite{freedman2007statistics}. 
Fig. \ref{fig:correlation_img} shows the results of the Pearson's correlation coefficient, where the PE is an extraversion marker of the IPIP Big-Five Factor Markers, the PES is emotional stability, the PA is agreeableness, the PC is conscientiousness, and the PII is an intellect/imagination. 

From the results of Pearson's correlation, we observed a moderate relationship between personal traits and physiological and behavioral features. The moderate positive relationship ($\gamma > 0.3$) is found between PE and mean IBI, between PES and mean GSR, SKT, and EAR, between PA and mean HR, and between the PC and mean IBI and EAR. On the other hand, the moderate negative relationship ($\gamma\textless-0.3$) is found between PE and mean HR, between PA and mean EAR and IBI, between PC and mean HR, and between PII and mean GSR, SKR, and EAR. 

\begin{table}[h]
\caption{Pearson's correlation coefficients between gender, personality traits, and performance measures.}
\label{tab:pearson_personality_perf}
\centering
\resizebox{\columnwidth}{!}{
\begin{tabular}{lccccc}
\hline \hline
                    & PE          & PA          & PES   & PC         & PI          \\ \hline
\textbf{All Participants} &             &             &       &            &             \\
Score               & .148$^{*}$  & -.124       & -.023 & .058       & -.128       \\
Success rate        & -.024       & -.153$^{*}$ & -.018 & -.044      & -.015       \\
Successful clicks   & .236$^{**}$ & -.034       & .001  & .105       & -.159$^{*}$ \\
Failure clicks      & .066        & .147$^{*}$  & .036  & .042       & -.002       \\ \hline
\textbf{Male}       &             &             &       &            &             \\
Score               & .204$^{*}$  & -.180       & -.025 & .031       & -.142       \\
Success rate        & .032        & -.103       & .015  & -.139      & -.124       \\
Successful clicks   & .269$^{**}$ & -.191$^{*}$ & -.014 & .184$^{*}$ & -.090       \\
Failure clicks      & .007        & .047        & .022  & .170       & .109        \\ \hline
\textbf{Female}     &             &             &       &            &             \\
Score               & .026        & -.044       & -.039 & .080       & -.116       \\
Success rate        & -.148       & -.234       & -.070 & .084       & .111        \\
Successful clicks   & .198        & .182        & -.002 & .005       & -.242$^{*}$ \\
Failure clicks      & .226        & .314$^{**}$ & .059  & -.119      & -.145       \\ \hline
\footnotesize{$^{**}$: $p$ \textless $.01$,  $^{*}$: $p$ \textless $.05$}
\end{tabular}}
\end{table}
\vspace{-10pt}

\begin{table}[]
\caption{Results of independent sample t-test of performance-related measures between male and female participants (df=185).}
\label{tab:independent-t}
\centering
\resizebox{\columnwidth}{!}{
\begin{tabular}{lccccccc}
\hline \hline
                   & \multicolumn{2}{c}{Male} & \multicolumn{2}{c}{Female} &       &      &           \\ \hline
                   & Mean       & SD          & Mean        & SD           & t     & p    & Cohen's d \\ \hline
Score              & 71.04      & 40.036      & 69.06       & 40.720       & .326  & .745 & .049      \\
Success rate       & .94        & .077        & .94         & .075         & -.246 & .806 & -.037     \\
Successful clicks  & 91.07      & 31.108      & 88.64       & 35.104       & .492  & .624 & .074      \\
Failure clicks & 6.68       & 8.833       & 6.53        & 8.696        & .111  & .912 & .017      \\ \hline
\end{tabular}}
\end{table}

\subsection{Relationships between personality traits, gender, and performance measures} 
We also utilized Pearson's correlation coefficient and independent sample t-test to examine the relationship between the participants' personal characteristics and performance results, such as scores and success rate. As shown in Table~\ref{tab:pearson_personality_perf}, participants with higher PE ratings demonstrated higher scores ($\gamma$=.148, $p$ \textless $.05$) and a higher number of successful clicks ($\gamma$=.236, $p$ \textless $.01$) during the experiment. Individuals with higher PA ratings had higher success rates in identifying abnormal objects. Furthermore, there was a positive correlation between participants' PI ratings and the number of successful clicks ($\gamma$=-.159, $p$ \textless $.05$).

The correlation coefficients between personality traits and performance, based on gender, are presented in Table~\ref{tab:pearson_personality_perf}, highlighting distinct gender characteristics during the experiment. Among male participants, higher PE scores were positively correlated with higher scores ($\gamma$=.204, $p$ \textless $.05$) and a higher frequency of successful clicks ($\gamma$=.269, $p$ \textless $.01$), while higher PC scores were positively associated with a higher number of successful clicks ($\gamma$=.184, $p$ \textless $.05$). Also, higher PA scores among male participants were negatively correlated with the number of successful clicks ($\gamma$=-.191, $p$ \textless $.05$). Among female participants, higher PA ratings showed a higher frequency of failure clicks ($\gamma$=.314, $p$ \textless $.01$), while higher PI scores had a negative correlation with the number of successful clicks ($\gamma$=-.242, $p$ \textless $.05$).

We conducted an independent sample t-test between male and female participants to examine the consistency of the observed tendencies in Table~\ref{tab:pearson_personality_perf}. The four performance measures demonstrated normality based on the Shapiro-Wilk test. As shown in Table~\ref{tab:independent-t}, there were no significant differences between genders in the performance measures, as supported by minimal effect sizes (Cohen
\'s d); (score: $t_{185}=.326$, $p=.745$, success rate: $t_{185}=-.246$, $p=.806$, number of successful clicks: $t_{185}=.492$, $p=.624$, and failure clicks: $t_{185}=.111$, $p=.912$). Therefore, we concluded that there were no significant differences in performance-related measures between genders.

\subsection{Subjective self-reporting annotations}
\begin{figure*}
    \centering
    \begin{subfigure}[b]{0.45\linewidth}
        \includegraphics[width=1\linewidth]{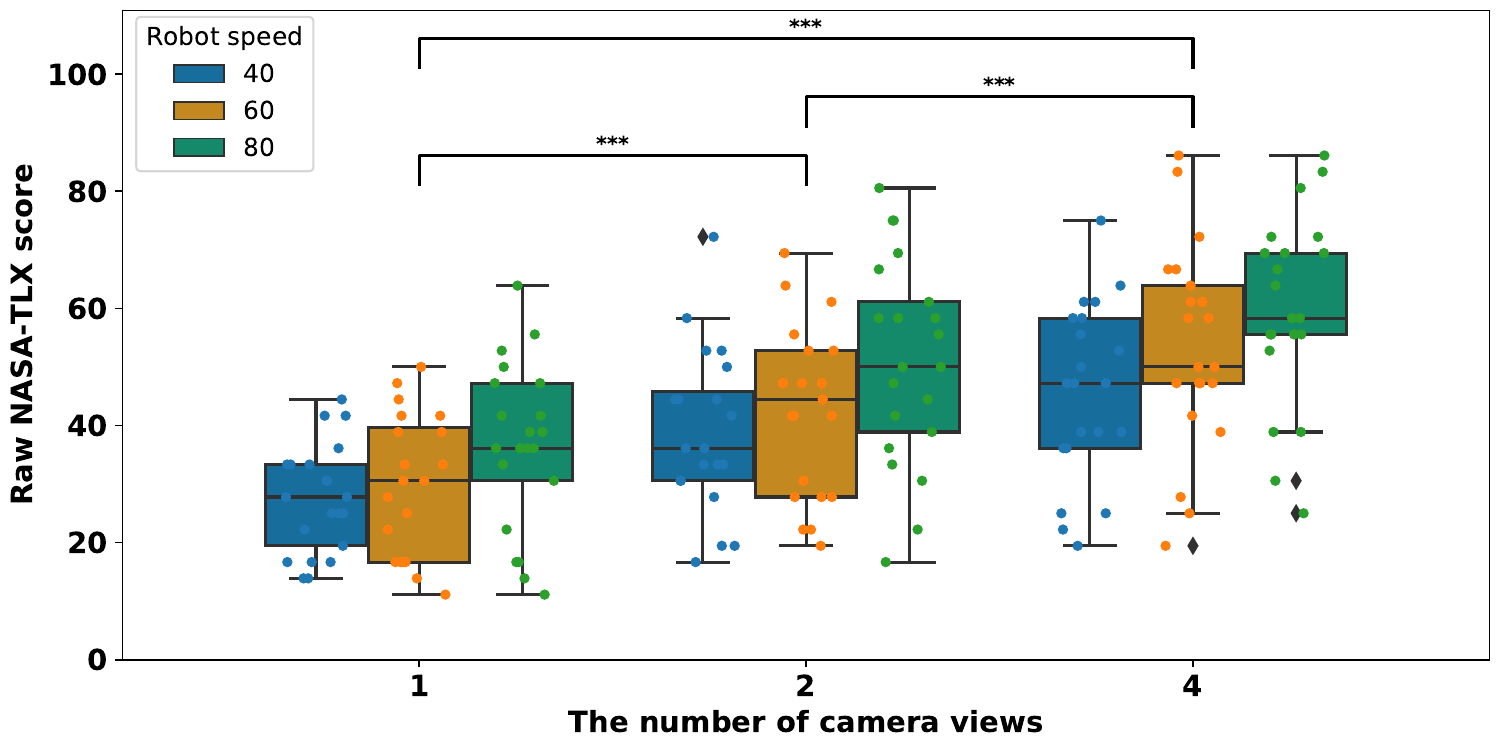}
        \caption{rmANOVA results of the raw NASA-TLX score}
        \label{img:anova_raw_nasa}
    \end{subfigure}
    \begin{subfigure}[b]{0.45\linewidth}
        \includegraphics[width=1\linewidth]{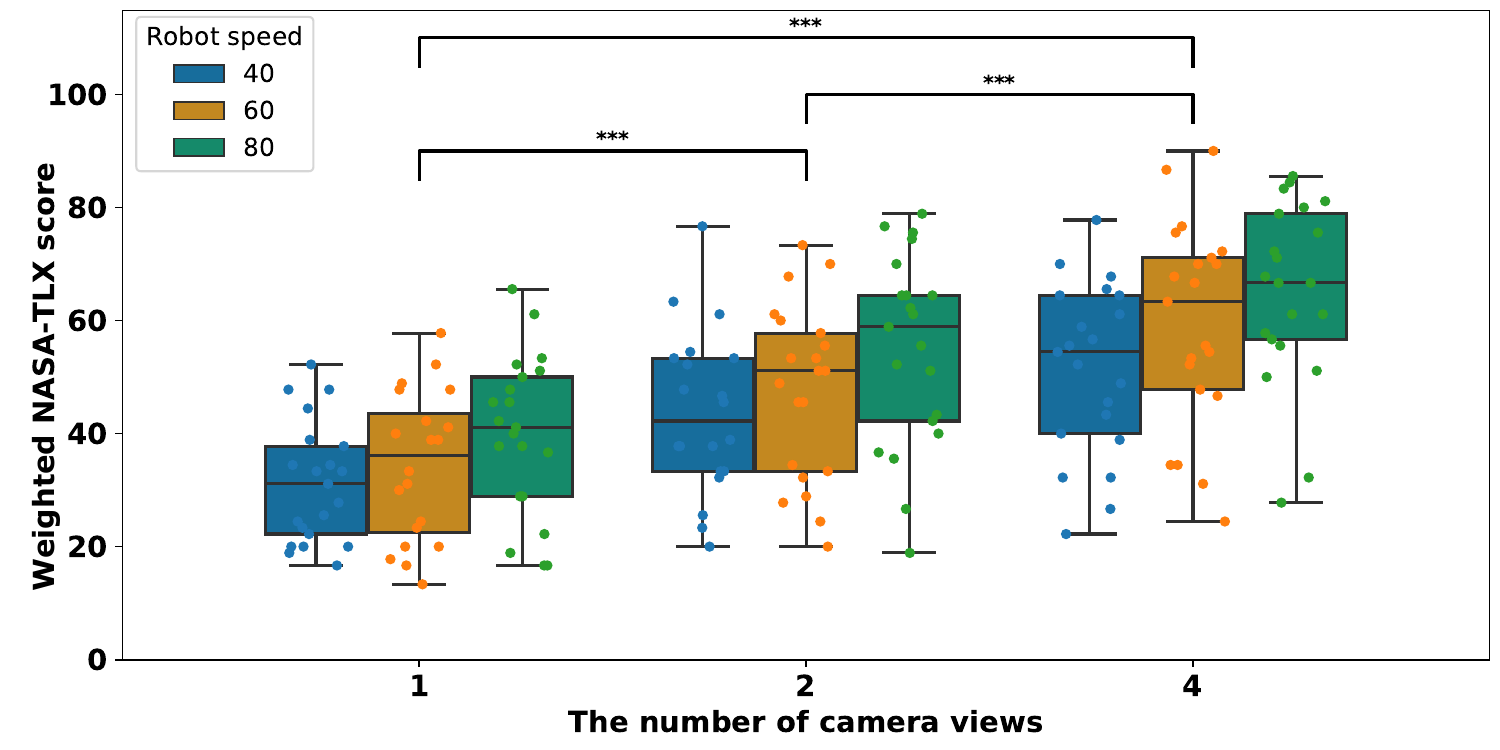}
        \caption{rmANOVA results of the weighted NASA-TLX score}
        \label{img:anova_weighted_nasa}
    \end{subfigure}
    \caption{The data distribution by variables of (a) raw NASA-TLX scores and (b) weighted NASA-TLX scores (***: $p\textless.001$).}
    \label{fig:cog_lod_results}
\end{figure*}

We conducted a two-way repeated-measures analysis of variance (rmANOVA) as well as a two-way Friedman test to validate the effects of the two within-subjects factors on participants' cognitive and emotional states. Initially, we carried out the Shapiro-Wilk test to examine whether the obtained subjective self-reporting responses fulfilled the normality assumption. The NASA-TLX and weighted NASA-TLX scores indicated normality, while the arousal and valence ratings in the SAM scale and ISA showed non-normal distribution. Therefore, we employed rmANOVA for the NASA-TLX and weighted NASA-TLX, and the Friedman test for SAM and ISA to demonstrated that the main task evoked distinct responses through the experiment conditions. In the rmANOVA test, the within-subjects factors were robot speed and the number of camera views (e.g., Robot\_speed and Camera\_number). We applied the Greenhouse-Geisser correction to address the violation of the sphericity assumption \cite{winer1971statistical}. The results of the rmANOVA, categorized by the type of subjective questionnaires, are summarized in Table \ref{tab:rmanova-nasa-gender}, and Fig. \ref{fig:cog_lod_results}. The two-way Friedman test does not require the assumptions of normality \cite{friedman1937use}. Therefore, we conducted a related samples Friedman's two-way analysis of variance by ranks for each self-questionnaire per task.

\begin{table*}[h]
\caption{Results of the two-way repeated-measures ANOVA test based on gender, raw and weighted NASA-TLX questionnaire self-reporting answers.}
\label{tab:rmanova-nasa-gender}
\centering
\resizebox{0.9\linewidth}{!}{
\begin{tabular}{ccccllll}
\hline \hline
\multicolumn{2}{c}{\textbf{Subjective questionnaires}}                  & \textbf{Factor}               & \textbf{$Df_{ef, err}$} & \multicolumn{1}{c}{\textbf{F}} & \multicolumn{1}{c}{\textbf{P}} & \multicolumn{1}{c}{\textbf{$\eta^{2}_{p}$}} & Note \\ \hline
\multicolumn{8}{l}{\textbf{All participants}} \\ \hline
\multirow{6}{*}{\textbf{NASA-TLX}} & \multirow{3}{*}{\textbf{Raw}}      & Camera\_number & (2, 38) & 49.127 & p \textless .001 & .721 & \\
& & Robot\_speed & (2, 38) & 21.030 & p \textless .001 & .525 & \\
& & Camera\_number x Robot\_speed & (4, 76) & 1.094 & .360 & .054 & \\ \cline{2-8} 
& \multirow{3}{*}{\textbf{Weighted}} & Camera\_number & (2, 38) & 59.303 & p \textless .001 & .757 & \\
& & Robot\_speed & (2, 38) & 17.303 & p \textless .001 & .477 & \\
& & Camera\_number x Robot\_speed & (4, 76) & .875 & .468 & .044 & \\ \hline

\multicolumn{8}{l}{\textbf{Male}} \\ \hline
\multirow{6}{*}{\textbf{NASA-TLX}} & \multirow{3}{*}{\textbf{Raw}} & Camera\_number & (2, 24) & 58.076 & p \textless .001 & .829 & \\
& & Robot\_speed & (2, 24) & 12.597 & .001 & .512 & ** \\
& & Camera\_number x Robot\_speed & (4, 48) & .474 & .695 & .038 & \\ \cline{2-8} 
& \multirow{3}{*}{\textbf{Weighted}} & Camera\_number & (2, 24) & 54.922 & p \textless .001 & .821 & \\
& & Robot\_speed & (2, 24) & 8.648 & .003 & .419 & ** \\
& & Camera\_number x Robot\_speed & (4, 48) & .362 & .770 & .029 & \\ \hline

\multicolumn{8}{l}{\textbf{Female}} \\ \hline
\multirow{6}{*}{\textbf{NASA-TLX}} & \multirow{3}{*}{\textbf{Raw}}      & Camera\_number   & (2, 12)  & 9.067  & .010  & .602  & **  \\
& & Robot\_speed   & (2, 12)  & 8.006  & .018  & .572  & **  \\
& & Camera\_number x Robot\_speed & (4, 24) & 1.048 & .373 & .149 & \\ \cline{2-8} 
& \multirow{3}{*}{\textbf{Weighted}} & Camera\_number & (2, 12) & 12.728 & .003 & .680 & **   \\
& & Robot\_speed & (2, 12) & 9.256 & .009 & .607 & **  \\
& & Camera\_number x Robot\_speed & (4, 24) & .931 & .402 & .134 &      \\ \hline
\footnotesize{$^{**}$: $p$ \textless $.05$}
\end{tabular}}
\end{table*}

\begin{table}[h]
\caption{Results of the two-way Friedman test on gender, ISA, and SAM.}
\label{tab:friedman-test}
\centering
\resizebox{0.8\columnwidth}{!}{
\begin{tabular}{cccc}
\hline \hline
\multicolumn{2}{c}{\textbf{Subjective questionnaires}} & \textbf{$\chi^2$} & \textbf{P} \\ \hline
\multicolumn{4}{l}{\textbf{All participants}} \\ \hline
\multicolumn{2}{c}{\textbf{ISA}} & 88.336 & p\textless{}.001 \\ 
\multirow{2}{*}{\textbf{SAM}} & \textbf{Arousal} & 49.161 & p\textless{}.001 \\ 
 & \textbf{Valence} & 63.282 & p\textless{}.001 \\ \hline
\multicolumn{4}{l}{\textbf{Male}} \\ \hline
\multicolumn{2}{c}{\textbf{ISA}} & 57.313 & p\textless{}.001 \\ 
\multirow{2}{*}{\textbf{SAM}} & \textbf{Arousal} & 35.113 & p\textless{}.001 \\ 
 & \textbf{Valence} & 45.116 & p\textless{}.001 \\ \hline
\multicolumn{4}{l}{\textbf{Female}} \\ \hline
\multicolumn{2}{c}{\textbf{ISA}} & 33.784 & p\textless{}.001 \\ 
\multirow{2}{*}{\textbf{SAM}} & \textbf{Arousal} & 18.719 & .016 \\ 
 & \textbf{Valence} & 21.553 & .006 \\ \hline
\end{tabular}}
\end{table}

\textbf{ISA results for measuring cognitive workload:} 
Table \ref{tab:rmanova-nasa-gender} presents the results of the two-way Friedman test for the ISA scores reported by participants. The scores were transformed into a range from -2 to 2, representing the string-type scores: Underutilized (-2), Relaxed (-1), Comfortable (0), High (1), and Excessive (2). The Friedman test revealed significant differences among repeated observations, yielding a Chi\--square ($\chi^2$) value of 88.336, with a significance level of $p\textless.001$. Further analysis of the ISA questionnaire was conducted separately for male and female participants, revealing noticeable differences. For male participants, the Friedman test showed a significant result ($\chi^2=57.313$, $p\textless.001$), while for female participants, the test yielded a significant result as well ($\chi^2=33.784$, $p\textless.001$). These findings suggest that the given task elicits distinct responses across different experimental conditions, regardless of the participants' gender.

\textbf{NASA-TLX results for measuring cognitive workload:} Fig. \ref{fig:cog_lod_results} illustrates the results of the two-way rmANOVA results of the raw and weighted NASA-TLX scores. The weights used in the weighted NASA-TLX score are $ [5, 0, 4, 3, 2, 1]$ in which the sequence is mental demand, physical demand, temporal demand, performance, effort, and frustration. From the two-way rmANOVA results using participants' raw and weighted NASA-TLX scores, we found a main effect of the number of the camera views (raw: $F_{2, 38}=49.12$, $p\textless.001$, $\eta_{p}^{2}=0.72$ and weighted: $F_{2, 38}=59.29$, $p\textless.001$, $\eta_{p}^{2}=0.76$) and the robot speed on the cognitive workload (raw: $F_{2, 38}=21.06$, $p\textless.001$, $\eta_{p}^{2}=0.53$ and weighted: $F_{2, 38}=17.36$, $p\textless.001$, $\eta_{p}^{2}=0.48$), but there is no interaction between the two within-subject factors (raw: $F_{4, 76}=1.09$, $p=0.37$, $\eta_{p}^{2}=0.05$ and weighted: $F_{4, 76}=0.86$, $p=0.49$, $\eta_{p}^{2}=0.04$). We also conducted a validation of the results based on gender to determine if there was congruence between male and female participants' responses. For male participants, both the raw and weighted NASA-TLX scores showed a significant main effect for two factors: the number of camera views (raw: $F_{2, 24}=58.076$, $p\textless.001$, $\eta_{p}^{2}=.829$; weighted: $F_{2, 24}=54.922$, $p\textless.001$, $\eta_{p}^{2}=.821$) and the robot speed (raw: $F_{2, 24}=12.597$, $p\textless.001$, $\eta_{p}^{2}=.512$; weighted: $F_{2, 24}=8.648$, $p=.003$, $\eta_{p}^{2}=0.419$). However, there was no interaction observed between these two factors (raw: $F_{4, 48}=.474$, $p=.695$, $\eta_{p}^{2}=.038$; weighted: $F_{4, 48}=.362$, $p=.770$. 
Likewise, the responses of female participants demonstrated congruence in the effects of the two factors. Significant main effects were observed for the number of cameras (raw: $F_{2, 12}=9.067$, $p=.010$ $(p\textless.05)$, $\eta_{p}^{2}=.602$; weighted: $F_{2, 12}=12.728$, $p=.003$ $(p\textless.05)$, $\eta_{p}^{2}=.680$) and the robot speed (raw: $F_{2, 12}=8.006$, $p=.018$, $\eta_{p}^{2}=.572$; weighted: $F_{2, 12}=9.256$, $p=.009$, $\eta_{p}^{2}=.607$) when analyzed separately. However, there was no significant interaction between these two factors (raw: $F_{4, 24}=1.048$, $p=.373$, $\eta_{p}^{2}=.149$; weighted: $F_{4, 24}=.931$, $p=.402$, $\eta_{p}^{2}=.134$). These results indicate that individual factors significantly influence cognitive workload, but the interaction between factors does not significantly impact participants' workload.

\textbf{SAM scores for measuring emotion state:}
Table \ref{tab:friedman-test}  
summarizes the results of the two-way Friedman test conducted on the SAM scores reported by the participants, using a 7-point Likert scale. From the test results, it was found that the main experiment significantly differentiated participants' responses in terms of arousal ($\chi^2=49.161$, $p\textless.001$) and valence ($\chi^2=63.282$, $p\textless.001$). These statistical findings hold true when the data are analyzed separately for each gender. Specifically, when considering arousal levels, both male participants ($\chi^2=35.113$, $p\textless.001$) and female participants ($\chi^2=18.719$, $p=.016$) exhibited statistically significant differences. Similarly, in terms of valence, both male participants ($\chi^2=45.116$, $p\textless.001$) and female participants ($\chi^2=21.553$, $p=.006$) showed statistically significant differences. These results indicate that the elicitation of emotions during the experiment is consistent across all genders.

\subsection{Classification Evaluation}  

\begin{figure}[t]
    \centering
    \includegraphics[width=1\linewidth]{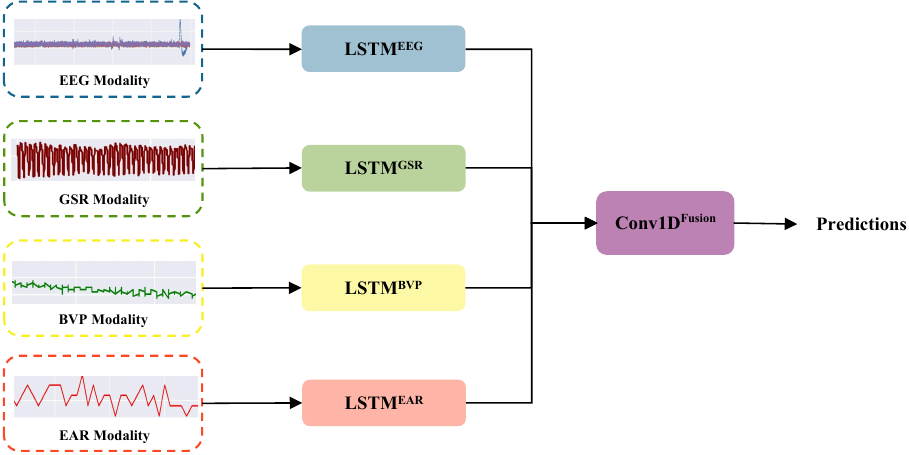}
    \caption{Illustration of LF-LSTM model for multimodal fusion classification using four example modalities as inputs}.
    \label{fig:lstm}
\end{figure}

\begin{table*}[t]
    \centering
    \caption{Evaluation results of unimodal and multimodal fusion three-class classification; Trial-Ind. denotes the trail-independent evaluation scheme, and Subject-Ind. presents the subject-independent, namely, LOSO, scheme.}
    \label{ACC}\resizebox{0.8\textwidth}{!}{%
    \begin{tabular}{ccccccc}
    \hline\hline
    \multicolumn{3}{c}{\textbf{Input Modality}} & \multicolumn{2}{c}{\textbf{Trial-Ind. Accuracy}} & \multicolumn{2}{c}{\textbf{Subject-Ind. Accuracy}} \\ \hline
    \textbf{Modality Name} & \textbf{Frequency} & \textbf{Channel Number} & \textbf{Mean (\%)}      & \textbf{Std}     & \textbf{Mean (\%)}      & \textbf{Std}     \\ \hline
    Fused                  & N/A                & N/A                     & 72.33              & 7.21             & 46.13     & 5.31            \\
    EEG                    & 128                & 6                       & 47.56              & 8.69             & 31.82              & 3.72            \\
    EEG\_POW               & 8                  & 25                      & 69.81              & 4.26             & 37.55              & 4.37            \\
    GSR                    & 4                  & 1                       & 40.52              & 5.68             & 28.37              & 5.15            \\
    BVP                    & 64                 & 1                       & 66.29              & 4.99             & 22.30             & 6.05            \\
    HR (from BVP)          & 1                  & 1                       & 68.84              & 5.92             & 23.22              & 4.69            \\
    IBI (from BVP)         & 1                  & 1                       & 60.18              & 4.15             & 27.09              & 3.56            \\
    EAR                    & 1                  & 3                       & 44.03              & 5.82             & 22.31             & 2.22            \\
    AU                     & 1                  & 3                       & 43.02              & 4.96             & 26.90              & 4.23            \\ \hline
    \end{tabular}
    }
\end{table*}

We conducted a comprehensive evaluation of the MOCAS dataset through a three-class cognitive workload classification. To this end, we utilized pre-processed features available in the pre-processed MOCAS dataset, which included EEG, EEG band powers (EEG\_POW), EDA (or GSR), BVP, HR, EAR and AU (see Section \ref{sec:preprocessed_down} and Table \ref{tab:data} for more information). In order to categorize the subjective cognitive workload annotations obtained from NASA-TLX questionnaires, we assigned them into three categories based on the literature \cite{grier2015high, favre2021high}. Specifically, scores ranging from 0-40 were labeled as ``low'' workload, scores from 40-60 as ``medium'' workload, and scores from 60-100 as ``high'' workload. Furthermore, we established and validated these score thresholds through multiple pilot tests, where the percentile statistics were empirically divided into three equal groups. The 33rd percentile was determined to be $37.78$ and the 66th percentile was found to be $55.56$. These values were then approximately rounded up to the previously mentioned ranges of 0-40, 40-60, and 60-100.

Taking inspiration from previous studies\cite{alhagry2017emotion,ma2019emotion} and recognizing the sequential nature of psychological and behavioral signals, we utilized a Long Short-Term Memory network (LSTM) \cite{graves2012long} for the task of unimodal classification. This approach involved using a single modality as the input. We also employed a Late-Fusion Long Short-Term Memory network (LF-LSTM) \cite{tsai2019multimodal,xu2018dual} for multimodal fusion classification, where multiple modalities were utilized as the input. As depicted in Fig. \ref{fig:lstm}, our LF-LSTM model processed the features from each modality through an LSTM network to extract time-related unimodal features. These processed unimodal features of all modalities were then concatenated and passed through a one-dimensional convolutional neural network (CNN) to generate predictions. For further details and access to the code of our LF-LSTM implementation, refer to the Code Availability section.

To emulate real-world application scenarios, we constructed input samples for each modality as 2-D feature matrices derived from one-second segments. Our evaluation methodology and data partitioning focused on two primary types: trial-independent and subject-independent schemes. In the trial-independent scheme, all samples were randomly shuffled, followed by $k$-fold cross-validation (where $k$=5). This process ensured that samples derived from the same monitoring task of a participant were distributed between the training and test sets. On the other hand, for the subject-independent scheme, we implemented the leave-one-subject-out (LOSO) cross-validation \cite{zhang2020emotion}.

The mean and standard deviation values of the three-class accuracy for both unimodal and multimodal fusion classifications under two evaluation schemes are detailed in Table \ref{ACC}. In the trial-independent evaluation, the highest achieved classification accuracy was $72.33\%$, reflecting the high quality of the data in the MOCAS dataset. Conversely, in the subject-independent evaluation, the highest accuracy decreased to $46.13\%$. This reduction is consistent with findings in previous studies \cite{miranda2018amigos,chen2021personal,zhang2020emotion} and can be attributed to the inherent challenges posed by the LOSO validation method. The LOSO approach, influenced by individual variances, complicates the development of a universally robust model. Moreover, our chosen classification network, the LF-LSTM, does not inherently adapt to these individual differences. Consequently, while the model demonstrates strong performance in a trial-independent context, its efficacy is diminished in the subject-independent scenario, reflecting the added complexity of generalizing across diverse individual datasets. This, in turn, provides an exciting venue for future research opportunities. The discrepancy in performance between trial-independent and subject-independent evaluations highlights the importance of developing more sophisticated models and techniques that can effectively account for individual differences. This area of research is particularly promising as it pushes the boundaries of current methodologies, encouraging innovation in personalized and adaptive modeling.

Subsequently, we conducted an independent two-sample t-test to compare the classification results across different modalities on the trial-independent evaluation. Generally, unimodal cognitive workload recognition using EEG\_POW modality performs significantly better than using EAR modality ($p=.05$), EEG modality ($p=.01$), and other single modalities ($p\textless.001$). EEG-related modalities, including EEG and EEG\_POW, perform well since EEG signals correspond more directly to different brain activities under different workload levels \cite{antonenko2010using}. Compared with EEG modality, EEG\_POW modality decomposes the raw EEG signals into component frequency bands, whose features are more intuitive and have fewer noises, leading to better classification results \cite{saby2012utility}. Moreover, EAR modality also achieves reasonable performance, we owe this to the fact that when facing different monitoring task levels where different numbers of camera views and robot speed, eye movements of participants would change correspondingly. Furthermore, classification using multimodal fusion significantly achieves better performance than using EEG\_POW ($p=.01$) and any other single modalities ($p\textless.001$). This reflects the benefits of the multimodal fusion mentioned in Section Background and Summary.

Furthermore, there is considerable scope for enhancing classification performance on the MOCAS dataset. Key areas for improvement include the adoption of more sophisticated classification models, as suggested by recent studies \cite{wang2022husformer,niu2020multimodal,zhou2020graph}, and refined data preprocessing techniques. Particularly for improving performance in subject-independent or LOSO evaluations, it is crucial to account for individual differences. Factors such as demographic information, personal traits, and task-specific experience should be considered. Advancements in transfer learning \cite{chen2021personal}, user-specific attention mechanisms \cite{liu2019nrpa,wang2023initial}, and few-shot learning \cite{feng2021few} methodologies are poised to make significant contributions in this regard. By integrating these approaches, we can develop more robust models that better accommodate the unique characteristics of individual subjects, thereby improving classification accuracy in diverse and personalized settings.

\section{Dataset access}
In order to protect the sensitive data and privacy of human subjects (e.g., physiological signals and facial views), only authorized researchers who consent to the End User License Agreement (EULA) are allowed to download the MOCAS. The researchers who want to access the MOCAS should visit our website and download the ELUA document; \url{https://polytechnic.purdue.edu/ahmrs/dataset}. After reviewing and filling the document up, they should email it to \url{info@smart-laboratory.org} and then request the access through Zenodo (\url{https://zenodo.org/}). Then, our research group will review and grant their access to our Zenodo repository having the downsampled MOCAS dataset, subjective information, and supplementary codes used in this paper. For sharing the raw dataset, we will sequentially invite their email address used in the Zenodo and the EULA document to access raw dataset uploaded on an additional repository (Purdue BOX, \url{https://purdue.box.com/v/mocas-dataset}), due to huge size of the raw dataset.

\subsection{Missing data}
Participant 2 (P2) discontinued experiments due to personal reasons, resulting in missing data for Task C and Task D from P2;
\begin{itemize}
        \item $P2\_mouse\_cam\_1\_speed\_60$
        \item $P2\_mouse\_cam\_2\_speed\_40$
\end{itemize}

Out of 187 files, one file does not contain the Emotive Insight data due to sensor disconnection. However, it includes other physiological and behavioral signals; 
\begin{itemize}
        \item $P3\_mouse\_cam\_4\_speed\_60\_signal\_0.db3$,
\end{itemize}

\noindent
where the term \textit{mouse} in the file name refers to a type of input interface used in user experiments. 

\subsection{Code availability}
\label{SECTION-CA}
The code utilized for data format converting (ROSbag2 to CSV) and preprocessing with the code of LSTM and LF-LSTM used for classification validation is included in the \textit{supplementary\_code} file of the dataset repository. Additionally, a command-line tool of the ROS2 middleware for replaying the ROSbag2 format files using the command of \textit{ros2 bag play \{rosbag2\_file\_name\}} or \textit{rqt\_bag} is also available there.

\section{Discussion \& Limitations}
We conducted experiments with real multi-robot systems to collect physiological and behavioral signals from human subjects while they watched real-time video streamed by the multi-robot systems. However, we encountered several challenges during the data collection process, including ensuring stable and real-time data collection, keeping the participants engaged, and dealing with discomfort caused by the EEG headset. To address these issues, we used ROS2 for real-time computing, designed realistic missions and provided extra compensation to increase participant engagement, and offered breaks during the experiment to alleviate the discomfort caused by the EEG headset.

Then, we built the MOCAS dataset by collecting behavioral and physiological signals using practical wearable biosensors and devices in real-life scenarios. Compared to other existing datasets listed in Table \ref{tab:T1}, our MOCAS dataset provides both raw physiological and behavioral data, as well as pre-processed features in real-time, from more realistic scenarios and stimuli. It also includes annotations of emotional states and cognitive load, which can be useful for training and evaluating machine learning models for different applications, such as emotion recognition and mental workload estimation. Moreover, the MOCAS dataset offers raw ROSbag files recorded during the entire experiment, enabling researchers to easily understand the situation and match it with the raw data. They can also test their prediction algorithms in real-time without requiring further experiments.

However, MOCAS dataset still has some limitations on the introduced dataset, so it should be considered before using this dataset as following limitations:

For the physiological signals, the commercial wearable biosensors used in this dataset tend to easily have unknown noises influenced by the participant's movement. There are available open-source libraries to remove common noises for the biosensors, such as \textit{NeuroKit2} \cite{Makowski2021neurokit}, \textit{pyphysio} \cite{bizzego2019pyphysio} and \textit{BioSPPy} \cite{biosppy2015}. The downsampled CSV files of our dataset were cleaned using the \textit{NeuroKit2} to remove the noises from the BVP and GSR signals.

For the behavioral data, the raw MOCAS dataset contains behavioral data obtained from the front facial videos of individuals who have agreed to share their data with the public. However, this dataset has a limitation in that it also includes behavioral data recorded from individuals wearing facial masks. This was due to the campus regulations during the data collection experiment, which mandated that everyone wear facial masks during the COVID-19 pandemic. Despite this limitation in the raw MOCAS, we extracted the behavioral features (such as EAR and AUs) using MediaPipe which can measure facial landmarks regardless of whether the individual is wearing a facial mask, then add the features on the prepossess MOCAS.

For the size of dataset and participants, this dataset comprises multimodal signals collected from 21 participants. We acknowledge that the number of subjects in our study is relatively small compared to certain existing dataset, with a total of 21 subjects included in the final analysis. As discussed in Section \ref{sec:dataset_sum}, the sample size was constrained by subject availability and the inclusion criteria for the study. Additionally, explicit informed consent was obtained from all participants, resulting in the exclusion of certain potential subjects from the final analysis.
In order to mitigate the limitations imposed by the dataset size, we employed $k$-fold cross-validation to construct a robust deep learning model for predicting human cognitive workloads. $k$-fold cross-validation is a widely recognized technique used to estimate the performance of a model on unseen data. It involves partitioning the dataset into $K$ equally sized folds and training the model $k$ times, with each fold serving as the test set once \cite{kohavi1995study}. This method is extensively employed in machine learning research and practice to mitigate overfitting and maximize the utilization of available data. By leveraging this validation approach, our deep learning model achieved a predictive accuracy of $74.68\%$ in categorizing three levels of cognitive workloads, namely low, medium, and high.

Other important limitation of this dataset is that it was collected from a restricted age range of participants. Although we aimed to select a diverse sample within this age range through official flyers and snowball sampling \cite{biernacki1981snowball}, our findings may not be generalizable to other age groups. This limitation is particularly relevant for phenomena known to vary across the lifespan, such as cognitive or physical abilities \cite{salthouse2019major}. Therefore, future studies could consider expanding the age range to include a wider range of ages, which would provide a more comprehensive understanding of the phenomenon under investigation \cite{bender2012normal}. In addition, future research could explore potential age-related differences in our variables of interest to further enhance our understanding of the phenomenon \cite{lovdn2020changes}. Despite this limitation, our study offers valuable insights within the age range studied and sets the stage for future research to build upon our findings.


\section{Conclusion} 
\label{conclusion}
We proposed a new multimodal dataset for human cognitive workload. The dataset includes physiological signals and behavioral features measured from 21 human subjects conducting the generalized CCTV monitoring task with a real multi-robot system. The physiological signals are acquired by the two wearable sensors, such as EEG, PPG, GSR, HR, IBI, SKT, and motion data. The behavioral features include eye ratios, facial expression, and facial action units extracted from a facial view of the webcam. The proposed dataset consists of raw and downsampled data, a summary of the participants' information, and the results of the subjective questionnaires. The total size of the raw dataset is about 722.4 GB, including 754 rosbag2 files. For the downsampled dataset, `\textit{.csv}' (Comma Separated Value) and `\textit{.pkl}' (Pickle) file formats were converted from raw datasets with sampling rates of 100 Hz. The size was 51.9 GB and 32.8 GB respectively. 

We also validated the quality of the dataset by analyzing the correlation between personality traits, physiological signals and behavioral features,  evaluating the effects of within-subjects factors on the results of the questionnaires presented to subjects, and applying an LF-LSTM to classify the three-class cognitive workload classifications. As a result, we found that there are significant differences in factors within subjects (e.g., the number of camera views and robot speed) using statistical analysis, and also showed that the classification performance of the multimodal dataset outperforms that of the single-modal dataset through deep learning methods.

Additionally, we made the MOCAS dataset publicly available by uploading the dataset to the online repositories and codes used in this dataset. We hope that the proposed dataset can become a fundamental resource for other researchers to develop systems and algorithms in human cognitive workloads. 

\section*{Acknowledgements} 
This material is based upon work supported by the National Science Foundation under Grant No. IIS-1846221. Any opinions, findings, and conclusions or recommendations expressed in this material are those of the author(s) and do not necessarily reflect the views of the National Science Foundation.

We also appreciate the valuable suggestions from anonymous reviewers and the editor regarding valuable comments in our study.

\bibliography{reference}
\bibliographystyle{IEEEtran}

\begin{IEEEbiography}[{\includegraphics[width=1in,height=1.25in,clip,keepaspectratio]{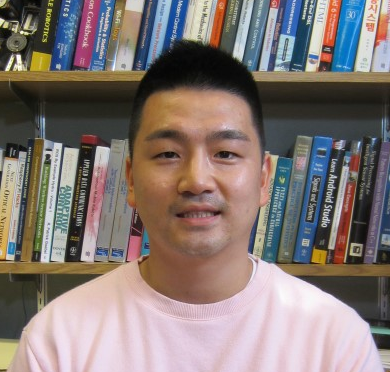}}]{Wonse Jo} received the B.S. in robotics engineering from Hoseo University, South Korea in 2013 and M.S. degrees in electronic engineering from the Kyung-Hee University, South Korea, in 2015. He is currently pursuing the Ph.D. degree in computer and information technology at Purdue University, West Lafayette, IN, USA. His research interests includes affective robotics/computing, human multi-robot interaction, environmental robotics, and field robotics
\end{IEEEbiography}
\begin{IEEEbiography}[{\includegraphics[width=1in,height=1.25in,clip,keepaspectratio]{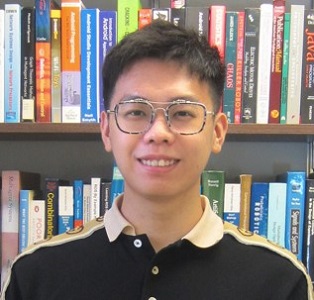}}]{Ruiqi Wang} (Student Member, IEEE) received a B.E. degree in robotics engineering from Beijing University of Chemical Technology, Beijing, China, in 2020. He is currently working towards a Ph.D. degree in the Department of Computer and Information Technology at Purdue University, West Lafayette, IN, USA. His research interests include human-robot interaction, affective robotics, and multi-modal deep learning.
\end{IEEEbiography}
\begin{IEEEbiography}[{\includegraphics[width=1in,height=1.25in,clip,keepaspectratio]{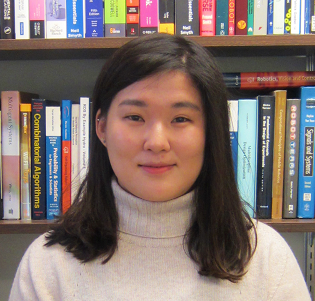}}]{Go-Eum Cha} received a B.S. degree in Computer Science and Engineering and B.E. degree in Global Media School from Soongsil University, Seoul, South Korea, in 2016, and M.S. degrees in Computer Information and Technology from Purdue University, USA. She is currently working towards a Ph.D. degree in the Department of Computer and Information Technology at Purdue University, West Lafayette, IN, USA. Her research interests include social robotics, cognitive and affective robotics/computing, and computer vision.
\end{IEEEbiography}
\begin{IEEEbiography}[{\includegraphics[width=1in,height=1.25in,clip,keepaspectratio]{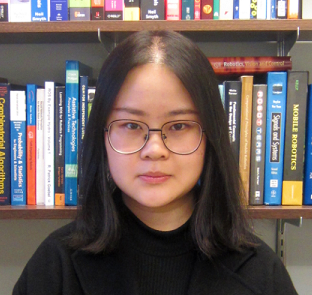}}]{Su Sun} received a B.E.degree in digital media technology from North China University of Technology, Beijing, China, in 2019. She is currently working towards a PhD degree in the Department of Computer and Information Technology at Purdue University, West Lafayette, IN USA. Her research interests include deep learning \& 3D geometry, SLAM, and 3D reconstruction.
\end{IEEEbiography}
\begin{IEEEbiography}[{\includegraphics[width=1in,height=1.25in,clip,keepaspectratio]{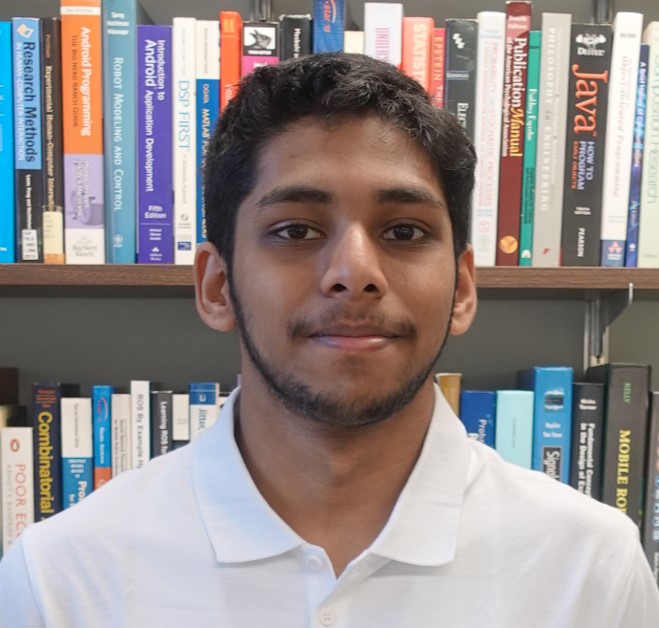}}]{Revanth Krishna Senthilkumaran} is currently an undergraduate student in the Department of Electrical and Computer Engineering at Purdue University, West Lafayette, IN, USA. His research interests include human-robot interaction, embedded electronics, robot dynamics, and unmanned aerial systems.
\end{IEEEbiography}
\begin{IEEEbiography}[{\includegraphics[width=1in,height=1.25in,clip,keepaspectratio]{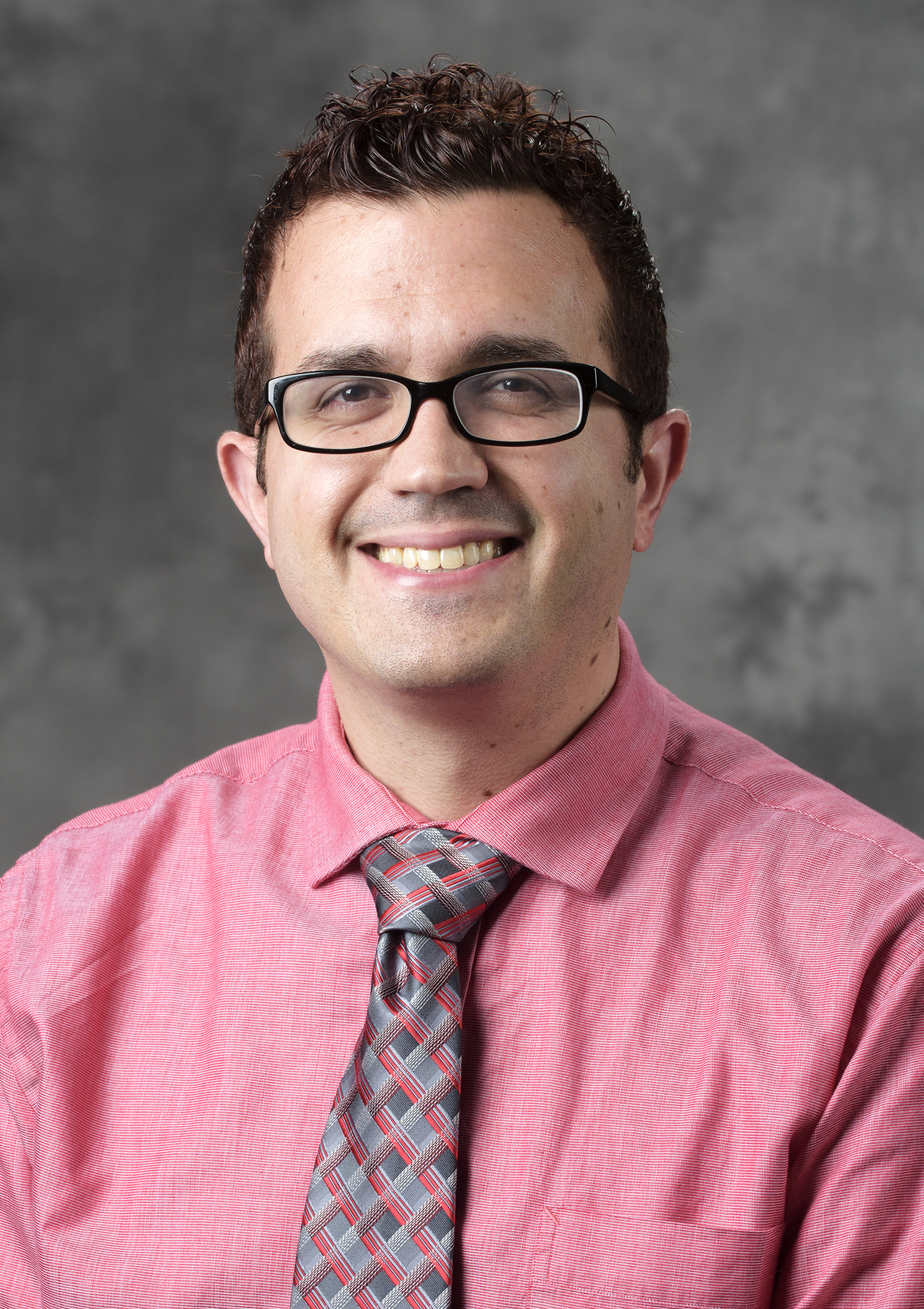}}]{Daniel Foti} received a B.A. degree in Biomedical Engineering from Harvard University in 2006. He completed his graduate studies at Stony Brook University, receiving his M.A. in Psychology in 2008 and his Ph.D. in Clinical Psychology in 2013. He completed his predoctoral clinical internship at McLean Hospital in Belmont, MA. He joined the faculty in the Department of Psychological Sciences at Purdue University as an Assistant Professor in 2013. He was promoted to Associate Professor with tenure in 2018. His research interests include using psychophysiological measures to study cognition, emotion, and reward, as well as abnormalities in these processes that are associated with vulnerability to psychiatric illnesses.

He was a recipient of the James C. Naylor Award for Teaching Excellence in Psychology in 2019. He was named Director of the T32 Predoctoral Training Program through the Indiana Clinical and Translational Sciences Institute in 2021. He was elected to the Board of Directors for the Society for Psychophysiological Research in 2021, and he was elected Treasurer in 2022.
\end{IEEEbiography}
\begin{IEEEbiography}[{\includegraphics[width=1in,height=1.25in,clip,keepaspectratio]{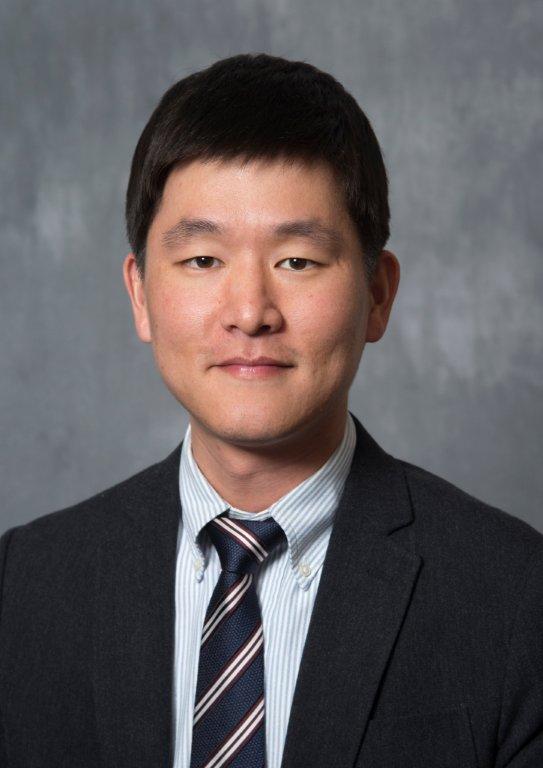}}]{Byung-Cheol Min} (Member, IEEE) received a B.S. degree in electronics engineering and a M.S. degree in electronics and radio engineering from Kyung Hee University, Yongin, South Korea, in 2008 and 2010, respectively, and a Ph.D. degree in technology with a specialization in robotics from Purdue University, West Lafayette, IN, USA, in 2014. He is currently an Associate Professor and University Faculty Scholar with the Department of Computer and Information Technology and the Director of the SMART Laboratory, Purdue University. Prior to this position, he was a Postdoctoral Fellow with the Robotics Institute, Carnegie Mellon University, Pittsburgh, PA, USA. His research interests include multi-robot systems, human–robot interaction, robot design and control, with applications in field robotics, and assistive technology and robotics.

He was a recipient of the NSF CAREER Award, in 2019; the Purdue PPI Outstanding Faculty in Discovery Award, in 2019; the Purdue CIT Outstanding Graduate Mentor Award, in 2019; the Purdue Focus Award, in 2020; the Purdue PPI Interdisciplinary Research Collaboration Award, in 2021; the Purdue Corps of Engagement Award, in 2022. He was named a Purdue University Faculty Scholar, in 2021.
\end{IEEEbiography}
\vspace{350pt}

\end{document}